\documentclass[journal]{IEEEtran}
\usepackage{cite}
\usepackage{amsmath,amssymb,amsfonts,dsfont, mathrsfs}
\usepackage[noend]{algpseudocode}
\usepackage{algorithmicx,algorithm}

\usepackage{graphicx}
\usepackage{subfigure}
\usepackage{textcomp}
\usepackage{xcolor}
\usepackage{booktabs}
\usepackage{breqn}
\usepackage{subeqnarray}
\usepackage{cases}
\usepackage{setspace}
\usepackage{amsthm}
\usepackage{bbm}
\usepackage{bm}
\usepackage{lipsum}
\usepackage{enumitem}
\usepackage{makecell}
\usepackage{multirow}
\usepackage{pifont}
\usepackage{epstopdf}
\usepackage{hyperref}

\allowdisplaybreaks[4]

\theoremstyle{remark}

\def\BibTeX{{\rm B\kern-.05em{\sc i\kern-.025em b}\kern-.08em
    T\kern-.1667em\lower.7ex\hbox{E}\kern-.125emX}}

\begin{document}

\title{Large Language Model-Empowered Channel Prediction and Predictive Beamforming for LEO Satellite Communications\\
}
\author{Zhixiong~Chen,~\IEEEmembership{Member,~IEEE},
Hyundong~Shin,~\IEEEmembership{Fellow,~IEEE}, \\
Arumugam~Nallanathan,~\IEEEmembership{Fellow,~IEEE}
and Jonathon~Chambers,~\IEEEmembership{Life Fellow,~IEEE}
\thanks{Zhixiong Chen and Arumugam Nallanathan are with the School of Electronic Engineering and Computer Science, Queen Mary University of London, E1 4NS London, U.K. (emails: zhixiong.chen@qmul.ac.uk, a.nallanathan@qmul.ac.uk).}
\thanks{Hyundong Shin is with the Department of Electronics and Information Convergence Engineering, Kyung Hee University, Yongin-si, Gyeonggido 17104, Republic of Korea (e-mail: hshin@khu.ac.kr).}
\thanks{Jonathon Chambers is with the School of Engineering, University of Leicester, Leicester, U.K. (e-mail: jac89@leicester.ac.uk).}
}

\maketitle
\begin{abstract}
Accurate channel prediction and effective beamforming are essential for low Earth orbit (LEO) satellite communications to enhance system capacity and enable high-speed connectivity. Most existing channel prediction and predictive beamforming methods are limited by model generalization capabilities and struggle to adapt to time-varying wireless propagation environments. Inspired by the remarkable generalization and reasoning capabilities of large language models (LLMs), this work proposes an LLM-based channel prediction framework, namely CPLLM, to forecast future channel state information (CSI) for LEO satellites based on historical CSI data. In the proposed CPLLM, a dedicated CSI encoder is designed to map raw CSI data into the textual embedding space, effectively bridging the modality gap and enabling the LLM to perform reliable reasoning over CSI data. Additionally, a CSI decoder is introduced to simultaneously predict CSI for multiple future time slots, substantially reducing the computational burden and inference latency associated with the inherent autoregressive decoding process of LLMs.
Then, instead of training the LLM from scratch, we adopt a parameter-efficient fine-tuning strategy, i.e., LoRA, for CPLLM, where the pretrained LLM remains frozen and trainable low-rank matrices are injected into each Transformer decoder layer to enable effective fine-tuning. Furthermore, we extend CPLLM to directly generate beamforming strategies for future time slots based on historical CSI data, namely BFLLM. This extended framework retains the same architecture as CPLLM, while introducing a dedicated beamforming decoder to output beamforming strategies. Finally, extensive simulation results validate the effectiveness of the proposed approaches in channel prediction and predictive beamforming for LEO satellite communications.
\end{abstract}

\section{Introduction}
The sixth generation (6G) of mobile communication envisions a hyper-connected world that supports ultra-reliable, low-latency, and high-throughput communication across diverse environments, including urban, rural, maritime, and airborne scenarios. To achieve these goals, low Earth orbit (LEO) satellite communication systems have garnered substantial attention as a key enabler of global connectivity \cite{10399870}. Operating at altitudes ranging from 500 to 2000 kilometers, LEO satellites offer significantly lower latency and higher link quality compared to medium and geostationary earth orbit satellites. As a result, LEO satellites are better suited to complement and extend terrestrial communication networks \cite{9982369}.

To improve transmission rates in LEO satellite communications, the multi-color reuse approach has been widely adopted, it assigns disjoint frequency bands to adjacent beams to mitigate co-channel inter-beam interference \cite{8654189}. Although the spectral efficiency can be improved by reusing different frequency bands across spatially isolated beams, multi-color reuse approaches face limitations in accommodating the growing number of user terminals and increasing service requirements. To address this challenge, multiple-input multiple-output (MIMO) transmission with full frequency reuse (FFR) was developed in \cite{9110855}, wherein all beams share the same frequency band to significantly enhance the spectral efficiency of LEO satellite communications. However, FFR inevitably introduces co-channel interference, necessitating advanced signal processing techniques for effective mitigation. Particularly, transmit-side beamforming can be employed to exploit spatial diversity and effectively suppress interference across beams.

Despite the effectiveness of transmit-side beamforming, it heavily relies on the timely and accurate acquisition of channel state information (CSI) \cite{10946972}.
However, due to long transmission delays and substantial Doppler shifts caused by the high-speed mobility of both LEO satellites and user terminals, the estimated CSI would be outdated before utilized for beamforming.
To address this challenge, various approaches were developed to predict forthcoming CSI based on historical CSI data. Specifically, the traditional model-based approaches employed parametric models such as autoregressive models \cite{1512123} and polynomial extrapolation models \cite{9127447} for channel prediction. However, their performance heavily depends on the accuracy of the assumed model, which usually struggles to capture the intricate characteristics of practical channels. With the impressive success of deep learning methods in diverse prediction tasks in various fields, deep learning-based CSI prediction has garnered substantial attention and has been extensively studied. In \cite{9210016}, a multilayer perceptron (MLP) model was trained to predict the forthcoming channels by capturing the characteristics of input historical CSI data.
Moreover, the recurrent neural network (RNN)-based channel predictors were developed in \cite{8813020, 8801923} to exploit the temporal dependencies of CSI over time, achieving superior performance compared to traditional model-based approaches. To overcome the vanishing gradient problem inherent in RNNs, a long short-term memory (LSTM) network-based channel prediction scheme was proposed in \cite{9439942} to predict channels in LEO satellite communication scenarios by leveraging the temporal correlations in channel variations. Hence, the gated recurrent unit (GRU)-based channel prediction method was developed in \cite{10089512} to simplify the model complexity of LSTM. In \cite{9921297}, a spatio-temporal neural network that combines LSTM and a convolutional neural network (CNN) was proposed to capture both the spatial correlations and temporal dependencies of CSI for channel prediction. Drawing inspiration from the powerful generalization capabilities of large language models (LLMs), the authors in \cite{10582829} proposed an LLM-based channel prediction approach for a single-user multiple-input single-output (MISO) system, achieving state-of-the-art prediction accuracy across various channel environments.

With the predicted CSI, beamforming techniques can be utilized to manage interference and optimize transmission performance for LEO satellite communications. In this context, commonly used methods in terrestrial communication scenarios, such as the zero-forcing \cite{4599181} and the weighted minimum mean squared error (WMMSE) algorithm \cite{4712693}, have been widely adopted in LEO satellite communications \cite{10440321, 9328170}. Considering that the LEO satellite is unlikely to obtain full CSI, a robust beamforming algorithm was proposed in \cite{9165811} to address channel phase uncertainty in two application scenarios. Moreover, the hybrid analog/digital precoding solution proposed in \cite{9694506} effectively addresses the high implementation complexity associated with fully digital precoding architectures in LEO satellite networks. A CNN-based hybrid precoder was proposed in \cite{9000850}, which effectively enhanced spectral efficiency in millimeter-wave MIMO communication systems. In \cite{10550141}, an LSTM-based channel prediction method was first introduced to forecast future CSI, followed by the use of a deep neural network to generate multi-beam precoding strategies based on the predicted CSI.

Nevertheless, the sequential process of first performing channel prediction followed by beamforming increases the signal processing overhead in LEO satellite communications. To address this problem, predictive beamforming has emerged as a promising solution, which directly predicts the beamforming strategy from historical CSI data. Specifically, an LSTM-based predictive beamforming scheme was developed in \cite{9143143} to  mitigate beam misalignment issues arising from unmanned aerial vehicle jittering. The convolutional LSTM-based predictive beamforming approach proposed in \cite{10138552} combines CNN and LSTM to extract spatial and temporal features from historical CSI, effectively enhancing beamforming performance. In \cite{10741218}, a Transformer-based predictive beamforming approach was developed to predict the precoding matrix based on historical CSI, which significantly improved data transmission rates in rate-splitting multiple access-enabled non-terrestrial networks.

While the aforementioned deep learning-based channel prediction and predictive beamforming methods have demonstrated satisfactory performance in terms of prediction accuracy and spectral efficiency, their capabilities are limited by the capacity of the underlying neural networks. Moreover, these approaches usually suffer from poor generalization in the highly dynamic and complex wireless propagation environments of LEO satellite communications.
Inspired by the impressive zero-shot and few-shot generalization capabilities of pretrained LLMs in cross-modality tasks, this work aims to empower frozen LLMs with the capability to perform channel prediction and predictive beamforming tasks for multi-user LEO satellite communications. Specifically, this work fine-tunes pretrained LLMs and proposes dedicated CSI data encoder and decoder modules, enabling the pretrained textual LLMs to better interpret and reason over wireless channel data. Unlike the LLM-based channel prediction approach in \cite{10582829}, which predicts the channel coefficients for each transmit-receive antenna pair individually, our proposed approaches simultaneously process the CSI data for all antenna pairs across all users. This not only accelerates the channel prediction process but also enables the LLM to effectively capture correlations across antennas, time, and users, thereby enhancing performance in both channel prediction and predictive beamforming.
The primary contributions of this work are listed as follows:
\begin{itemize}
  \item We propose an LLM-based channel prediction framework, namely CPLLM, which leverages the reasoning and generalization capabilities of pretrained LLMs to forecast future CSI from historical CSI data, enabling LEO satellites to acquire timely and accurate downlink CSI. Within this framework, a preprocessing module is proposed to convert the CSI data to meet the required format of neural networks. Then, a dedicated CSI encoder module is designed to transform CSI data into embedded token vectors, allowing the LLM to interpret and capture the temporal and spatial correlations inherent in the channel. Additionally, a CSI decoder is introduced to predict CSI for multiple future time slots simultaneously, effectively reducing the computational costs and inference latency associated with conventional sequential autoregressive decoding strategies.

  \item Instead of training CPLLM from scratch, we adopt a parameter-efficient fine-tuning strategy, i.e., LoRA, in which the pretrained LLM is kept frozen while trainable low-rank matrices are injected into each Transformer decoder layer of the LLM. During the fine-tuning process, only the parameters of the CSI encoder, LoRA adapters, and CSI decoder are updated, accounting for only a small fraction of the total model parameters. This makes the fine-tuning process both highly efficient and cost-effective.

  \item To reduce the signal processing overhead and error propagation associated with the sequential process of first performing channel prediction and then beamforming, we extend CPLLM to directly predict beamforming strategies for future time slots using historical CSI data, namely BFLLM. This extended framework shares the same architecture as the former CPLLM framework in the data preprocessing module, CSI encoder and LLM backbone, while incorporating a dedicated beamforming decoder designed to output the corresponding beamforming strategies.

  \item We conduct extensive experiments to verify the effectiveness of the proposed CPLLM and BFLLM. The results demonstrate that our methods outperform all benchmark schemes in both channel prediction accuracy and beamforming performance. Specifically, compared to the state-of-the-art channel prediction approach in \cite{10582829}, the proposed CPLLM achieves a 6 dB reduction in normalized mean square error (NMSE) for channel prediction. In addition, in comparison with the benchmark predictive beamforming strategy, our proposed BFLLM boosts the sum rate of LEO satellite downlink transmissions by 36\%.
\end{itemize}

The remainder of this paper is structured as:
In Section \ref{sec:system_model}, we present the channel and signal models for the LEO satellite communication system.
Section \ref{sec:channel_prediction} formulates the channel prediction problem and elaborates on the proposed CPLLM.
In Section \ref{sec:beamforming}, we extend the proposed CPLLM to directly predict the beamforming strategies based on historical CSI data.
Simulation results are presented in Section \ref{sec:simulations} to verify the effectiveness of our proposed approaches.
Finally, we conclude this work in Section \ref{sec:conclusion}.

\section{System Model}\label{sec:system_model}
\begin{figure}[t]
\centering
\includegraphics[width=0.42\textwidth]{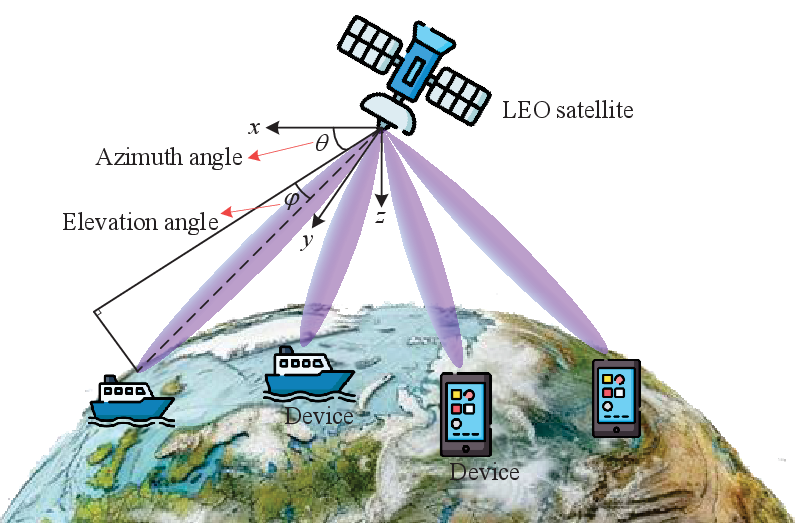}\\
\caption{The illustrated LEO satellite communication system.}
\label{fig:sys_model}
\end{figure}

In this work, we consider the downlink data transmission scenario of an LEO satellite communication system, as shown in Fig. \ref{fig:sys_model}. The LEO satellite is equipped with a uniform planar array (UPA) comprising $N = N_x \times N_y$ antennas to simultaneously communicate with $K$ widely distributed single-antenna users within its service area, where $N_x$ and $N_y$  represent the number of antennas along the $x$-axis and $y$-axis, respectively.
The devices are indexed by $\mathcal{K} = \{1, 2, \cdots, K\}$.
The system operates in a time-slotted manner, with each slot comprising two phases: one for CSI acquisition and another for downlink transmission.
The CSI acquisition process involves channel estimation at the devices followed by feedback of the estimated CSI to the LEO satellite.
Due to the high mobility of both the LEO satellite and user devices, the channel coherence time is extremely short. Additionally, the long transmission distance between the LEO satellite and devices introduces considerable feedback latency. As a result, acquiring timely downlink CSI becomes highly challenging.
To this end, this work focuses on predicting future CSI and beamforming strategies based on historical CSI.
In the following, we first present the channel model, followed by the downlink signal transmission model.

\subsection{Channel Model}
Based on the signal propagation characteristics of the LEO satellite communication system \cite{6395846, 9439942, 10742081}, the downlink channel from the LEO satellite to the $k$-th ($\forall k \in \mathcal{K}$) device at time $t$ over carrier frequency $f$ can be described as follows:
\begin{align}
\mathbf{h}_k(t,f) = \sqrt{\frac{\kappa_k}{\kappa_k \!+\! 1}} \mathbf{h}_k^{\rm{LOS}}(t,f) + \sqrt{\frac{1}{\kappa_k \!+\! 1}} \mathbf{h}_k^{\rm{NLOS}}(t,f).
\end{align}
Here, $\kappa_k$ denotes the Rician factor, while $\mathbf{h}_k^{\rm{LOS}}(t,f) \in \mathbb{C}^{N \times 1}$ and $\mathbf{h}_k^{\rm{NLOS}}(t,f) \in \mathbb{C}^{N \times 1}$ represent the line-of-sight (LOS) and non-line-of-sight (NLOS) components of the channel, respectively.
Considering that scatterers remain stationary within the interval of interest,
$\mathbf{h}_k^{\rm{LOS}}(t,f)$ is given by
\begin{align}
\mathbf{h}_k^{\rm{LOS}}(t,f) = & \alpha_k^{\rm{LOS}} e^{j2\pi [ t(  f_{{\rm{d}},k}^{\rm{Sat},\rm{LOS}} + f_{{\rm{d}},k}^{\rm{Dev},\rm{LOS}} ) - f \tau_k^{\rm{LOS}} ]} \notag\\
&\cdot \mathbf{u}_k\left(\theta_k^{\rm{LOS}},\varphi_k^{\rm{LOS}} \right),
\end{align}
and $\mathbf{h}_k^{\rm{NLOS}}(t,f)$ is given by
\begin{align}
\mathbf{h}_k^{\rm{NLOS}}(t,f) = & \sqrt \frac{1}{L_k} \sum\limits_{l = 1}^{L_k} \alpha_{k,l}  \cdot e^{j2\pi t\left(f_{{\rm{d}},k,l}^{\rm{Sat},\rm{NLOS}} + f_{{\rm{d}},k,l}^{\rm{Dev},\rm{NLOS}} \right)} \notag\\
& \cdot e^{-j2\pi f \tau_{k,l}^{\rm{NLOS}}} \cdot \mathbf{u}_k\left(\theta_{k,l}^{\rm{NLOS}}, \varphi_{k,l}^{\rm{NLOS}} \right),
\end{align}
where $L_k$ denotes the number of NLOS propagation paths of device $k$. $\alpha_k^{\rm{LOS}}$ and $\alpha_{k, l}$ are the complex-valued gain of the LOS path and the $l$-th NLOS path, respectively.
$f_{{\rm{d}},k}^{\rm{Sat},\rm{LOS}}$ ($f_{{\rm{d}},k,l}^{\rm{Sat},\rm{NLOS}}$ ) and $f_{{\rm{d}},k}^{\rm{Dev},\rm{LOS}}$ ($ f_{{\rm{d}},k,l}^{\rm{Dev},\rm{NLOS}}$) are the Doppler shift caused by the motion of satellite and device $k$ at the LOS path (the $l$-th NLOS path) of device $k$, respectively.
$\tau_k^{\rm{LOS}}$ and $\tau_{k,l}^{\rm{NLOS}}$ represent the propagation delay of device $k$ at the LOS path and $l$-th NLOS path, respectively.
$\theta_k^{\rm{LOS}}$ ($\varphi_k^{\rm{LOS}}$ ) and $\theta_{k,l}^{\rm{NLOS}}$ ($\varphi_{k,l}^{\rm{NLOS}}$) are the angle of horizontal (vertical) directions associated with the LOS and $l$-th NLOS path of device $k$, respectively.
$\mathbf{u}_k(\theta, \varphi)$ ($\theta \in \{\theta_{k}^{\rm{LOS}},\theta_{k,l}^{\rm{NLOS}}\}$, $\varphi \in \{ \varphi_{k}^{\rm{LOS}}, \varphi_{k,l}^{\rm{NLOS}} \}$) is the array response vector and can be expressed as:
\begin{align}
\mathbf{u}_k(\theta ,\varphi) = & \frac{1}{\sqrt N} \left[ e^{ - j2\pi d\sin \theta \sin \varphi \mathbf{n}_x/\lambda } \right] \notag \\
&\otimes \left[ e^{ - j2\pi d\cos \varphi \mathbf{n}_y/\lambda } \right],
\end{align}
where $\mathbf{n}_x = [0, 1, \cdots, N_x-1]^T$ and $\mathbf{n}_y = [0, 1, \cdots, N_y-1]^T$.
$\lambda$ is the wavelength of the carrier, $d$ is the antenna spacing and usually set as one-half wavelength, $c$ is the speed of light, and $\otimes$ is the Kronecker product.

Unlike terrestrial communication scenarios, satellite communications encounter significant Doppler shifts and transmission delays due to the satellite's high velocity and extended propagation distances.
The Doppler shift for the LOS path (or any NLOS path $l$) of device $k$ is induced by the movements of both the LEO satellite and the device, i.e., $f_{{\rm{d}},k}^{\rm{Sat},\rm{LOS}}$ ($f_{{\rm{d}},k,l}^{\rm{Sat},\rm{NLOS}}$ ) and $f_{{\rm{d}},k}^{\rm{Dev},\rm{LOS}}$ ($ f_{{\rm{d}},k,l}^{\rm{Dev},\rm{NLOS}}$).
Due to the high altitude of the LEO satellite, the relative motion between the satellite and a device is similar across various propagation paths. Consequently, the Doppler shift induced by satellite movement can be considered nearly identical across different paths \cite{662636}, i.e.,  $f_{{\rm{d}},k}^{\rm{Sat},\rm{LOS}} \approx f_{{\rm{d}},k,l}^{\rm{Sat},\rm{NLOS}}$.
In contrast, the Doppler shift caused by device movement typically varies across different propagation paths, resembling the observations in terrestrial networks.
Furthermore, we define the delay for the $l$-th NLOS path of device $k$ as $\Delta \tau_{k,l} = \tau_{k,l}^{\rm{NLOS}} - \tau_{k}^{\rm{LOS}}$, then $\mathbf{h}_k^{\rm{NLOS}}(t,f)$ can be rewritten as
\begin{align}
\mathbf{h}_k^{\rm{NLOS}}(t,f) &=  \sqrt \frac{1}{L_k} \sum\limits_{l = 1}^{L_k} \alpha_{k,l}  \cdot e^{j2\pi t\left(f_{{\rm{d}},k}^{\rm{Sat},\rm{LOS}} + f_{{\rm{d}},k,l}^{\rm{Dev},\rm{NLOS}} \right)} \notag\\
& \cdot e^{-j2\pi f (\Delta \tau_{k,l} + \tau_{k}^{\rm{LOS}})} \cdot \mathbf{u}_k\left(\theta_{k,l}^{\rm{NLOS}}, \varphi_{k,l}^{\rm{NLOS}} \right).
\end{align}

\subsection{Signal Model}
Let $\mathbf{s}_t = [s_{1, t}, s_{2,t}, \cdots, s_{K, t}]$ denote the transmitted data from the LEO satellite to devices in time slot $t$, where $s_{k,t}$ represents the data signal with unit-power intended for device $k$.
In the downlink transmission, the LEO satellite constructs a superposition-coded signal $\mathbf{x}_t$ by summing the products of each $s_{k,t}$ and its corresponding beamforming vector, i.e.,
\begin{align}
\mathbf{x}_t = \sum\nolimits_{k = 1}^K \mathbf{w}_{k,t} s_{k,t},
\end{align}
where $\mathbf{w}_{k,t} \in \mathbb{C}^{N\times 1}$ is the beamforming vector for device $k$. Then, the LEO satellite transmits $\mathbf{x}_t$ to devices through the downlink channels. The received signal at device $k$ is
\begin{align}\label{eq:received_signal}
&y_{k}(t,f) = \mathbf{h}_k^H(t, f) \mathbf{x}_t + n_{k,t}  \notag\\
&= \mathbf{h}_k^H(t, f) \mathbf{w}_{k,t} s_{k,t} + \!\!\!\!\sum\limits_{j = 1,j \ne k}^K \!\!\! \mathbf{h}_k^H(t, f) \mathbf{w}_{j,t} s_{j,t}  + n_{k}(t,f),
\end{align}
where the first term on the right-hand side of \eqref{eq:received_signal} represents the desired signal, while the second term denotes inter-device interference. $n_{k}(t,f)$ is the additive white Gaussian noise with zero mean and variance $\sigma^2$.
Hence, the signal to inference plus noise ratio (SINR) of device $k$ is given by
\begin{align}\label{eq:interference_k}
\Gamma_{k}(t,f) = \frac{\left| \mathbf{h}_k^H(t, f) \mathbf{w}_{k,t} \right|^2}  {\sum\nolimits_{j = 1,j \ne k}^K \left| \mathbf{h}_k^H(t, f) \mathbf{w}_{j,t} \right|^2 + \sigma^2 }.
\end{align}
Consequently, the achievable transmission rate of device $k$ can be described as
\begin{align}\label{eq:rate_k}
R_k(t,f) = \log_2\left(1 + \Gamma_{k}(t,f)\right).
\end{align}

As shown in \eqref{eq:interference_k} and \eqref{eq:rate_k}, the beamforming vectors, i.e., $\mathbf{w}_{k,t}$ ($\forall k \in \mathcal{K}$), significantly influence the achievable transmission rates for devices.
Thus, it is necessary to judiciously optimize the beamforming strategies to enhance the overall downlink transmission performance of the LEO satellite communication system.
To this end, we aim to maximize the sum rate of all devices in time slot $t$ by optimizing their beamforming strategies. Specifically, the optimization problem is formulated as follows:
\begin{align}
\max_{\left\{ \mathbf{w}_{k,t} \right\}_{k = 1}^{K}}&~ \sum\nolimits_{k = 1}^K \log_2(1 + \Gamma_{k}(t,f))\label{prob:P}\\
\text{s.~t.~~} & \sum\nolimits_{k = 1}^K \left\| \mathbf{w}_{k,t} \right\|^2 \le P_T, \label{cons:P_1}\tag{\theequation a}
\end{align}
where $P_T$ is the maximum transmit power of the LEO satellite.

Problem \eqref{prob:P} is a typical non-convex optimization problem, and its solution requires accurate downlink CSI from the LEO satellite to all devices.
With available CSI at slot $t$, i.e., $\{\mathbf{h}_k(t,f): \forall k \in \mathcal{K}\}$, existing beamforming methods, such as WMMSE \cite{4712693}, can be utilized to find a suboptimal solution for problem \eqref{prob:P}.
However, due to the high mobility of the LEO satellite and devices, obtaining timely downlink CSI is challenging, potentially leading to degraded beamforming performance and reduced downlink transmission efficiency.
To address this challenge, this work proposes an LLM-based channel prediction approach, i.e., CPLLM, in Section \ref{sec:channel_prediction}, wherein the LEO satellite predicts downlink CSI for future time slots using historical CSI data. Based on the predicted CSI, the satellite can then solve problem \eqref{prob:P} to generate downlink beamforming strategies.
Furthermore, considering the high computational overhead associated with solving beamforming problems, we extend the proposed CPLLM to directly predict beamforming strategies for future slots using CSI from past slots in Section \ref{sec:beamforming}.

\begin{figure*}[t]
\centering
\includegraphics[width=0.8\textwidth]{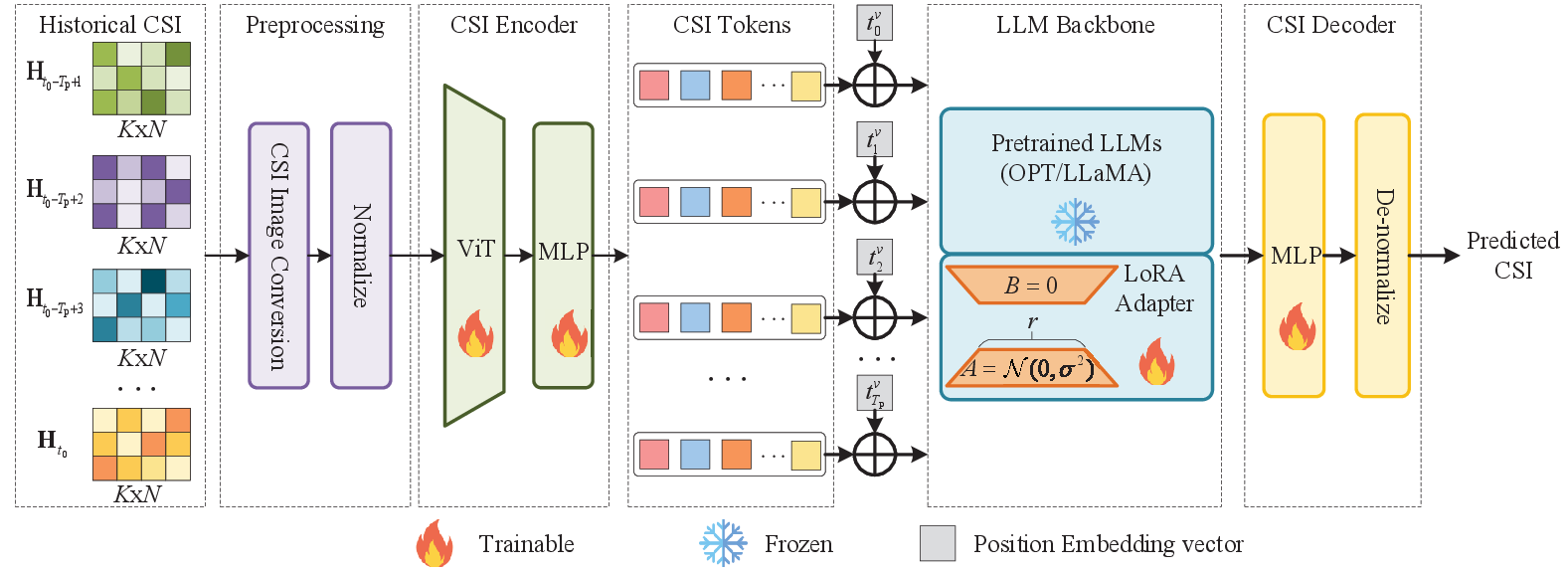}\\
\caption{The architecture of the proposed LLM-based channel predictor.}
\label{fig:LLM_channel_predictor}
\end{figure*}

\section{LLM-Based Channel Prediction}\label{sec:channel_prediction}
This section illustrates the proposed CPLLM, which empowers frozen LLMs with the capability to capture the spatial and temporal correlations of downlink CSI across adjacent time slots, enabling accurate prediction of future CSI based on the estimated CSI in previous time slots.

\subsection{Channel Prediction Problem Formulation}
In this work, we utilize the historical downlink CSI of devices over the past $T_{\rm{P}}$ time slots to predict the CSI for the upcoming $T_{\rm{F}}$ time slots.
For clarity, we denote the estimated CSI of all devices at time slot $t$ by $\mathbf{H}_t = [\mathbf{h}_1(t, f), \mathbf{h}_2(t, f), \cdots, \mathbf{h}_K(t, f)]^T \in \mathbb{C}^{K\times N}$.
The objective of channel prediction is to minimize the normalized mean square error (NMSE) between the predicted and the actual CSI in the future $T_{\rm{F}}$ time slots.
Accordingly, we formulate the channel prediction problem as follows:
\begin{align}
\min_{\bm{\Omega}} & ~  \mathbb{E} \Bigg[ \frac{\sum\nolimits_{t = 1}^{T_{\rm{F}}} {\left\| \mathbf{H}_{t_0+t} - \widehat{\mathbf{H}}_{t_0+t} \right\|^2}} {\sum\nolimits_{t = 1}^{T_{\rm{F}}} \left\| \mathbf{H}_{t_0+t} \right\|^2 } \Bigg] \label{prob:channel_prediction}\\
\text{s.~t.} ~& \big[ \widehat{\mathbf{H}}_{t_0+1}, \cdots , \widehat{\mathbf{H}}_{t_0 + T_{\rm{F}}} \big] = f_{\bm{\Omega}} (\left[ \mathbf{H}_{t_0-T_{\rm{P}}+1}, \cdots, \mathbf{H}_{t_0} \right]), \tag{\theequation a}
\end{align}
where $f_{\bm{\Omega}}(\cdot)$ is the proposed CPLLM,  $\bm{\Omega}$ is the set of model parameters. $\widehat{\mathbf{H}}_{t}$ is the predicted CSI in time slot $t$.

\subsection{LLM-Based Channel Predictor Framework}\label{subsec:CP_LLM}
In this subsection, we develop an LLM-based channel predictor, i.e., CPLLM, to solve problem \eqref{prob:channel_prediction}.
The overall architecture of CPLLM is depicted in Fig. \ref{fig:LLM_channel_predictor}. It consists of a CSI preprocessing module, a CSI encoder, an LLM backbone, and a CSI decoder. In the following, we provide detailed illustrations of each component.

1) \textbf{CSI preprocessing module}:
During the channel prediction process, the historical CSI data over the past $T_{\rm{P}}$ time slots, i.e., $[\mathbf{H}_{t_0-T_{\rm{P}}+1}, \cdots, \mathbf{H}_{t_0}]$, are fed into CPLLM to forecast the CSI for the upcoming $T_{\rm{F}}$ time slots.
Since neural networks operate on real-valued data, while CSI is inherently complex-valued, a transformation is necessary to make the data compatible with neural network processing.
To this end, each CSI matrix $\mathbf{H}_{t}$ ($t \in \{t_0-T_{\rm{P}}+1, \cdots, t_0\}$) is converted into a two-channel 2D image $\mathbf{H}_t^{\rm{real}}$, referred to as a CSI image, with one channel representing the real part and another representing the imaginary part. Formally, $\mathbf{H}_t^{\rm{real}}$ is expressed as
\begin{align}
\mathbf{H}_t^{\rm{real}} = \left[ \mathfrak{R}(\mathbf{H}_t),\mathfrak{I}(\mathbf{H}_t) \right] \in \mathbb{R}^{2 \times K \times N},
\end{align}
where $\mathfrak{R}(\mathbf{H}_t)$ and $\mathfrak{I}(\mathbf{H}_t)$ denote the real and imaginary components of $\mathbf{H}_t$, respectively.
Moreover, to facilitate the training and convergence of CPLLM, each CSI image is normalized as $\mathbf{X}_t^{\rm{pre}} =\frac{\mathbf{H}_t^{\rm{real}}- \mu_{\rm{b}}} {\sigma_{\rm{b}}}$, where $\mu_{\rm{b}}$ and $\sigma_{\rm{b}}$ denote the mean and standard deviation of a batch of input CSI images, respectively.

2) \textbf{CSI encoder module}:
After preprocessing, we obtain $T_{\rm{P}}$ normalized CSI images, i.e., $[\mathbf{X}_t^{\rm{pre}}: t \in \{t_0-T_{\rm{P}}+1, \cdots, t_0\} ]$, each of which corresponding to the CSI matrix from one of the past $T_{\rm{P}}$ time slots.
Since LLMs are primarily designed to process textual data, which differs significantly from CSI data, directly feeding these CSI images into LLMs would be ineffective and likely result in poor performance.
To address this issue, we propose a CSI encoder module to map each CSI image $\mathbf{X}_t^{\rm{pre}} \in \mathbb{R}^{2 \times K \times N}$ into an embedding vector $\mathbf{x}_t^{\rm{emb}} \in \mathbb{R}^{1 \times d_{\rm{LLM}}}$, yielding CSI representations $\mathbf{X}^{\rm{emb}} = [\mathbf{x}_{t_0-T_{\rm{P}}+1}^{\rm{emb}}, \cdots, \mathbf{x}_{t_0}^{\rm{emb}}]$, where $d_{\rm{LLM}}$ is the size of the embedding vectors in LLMs.

\begin{figure}[t]
\centering
\includegraphics[width=0.5\textwidth]{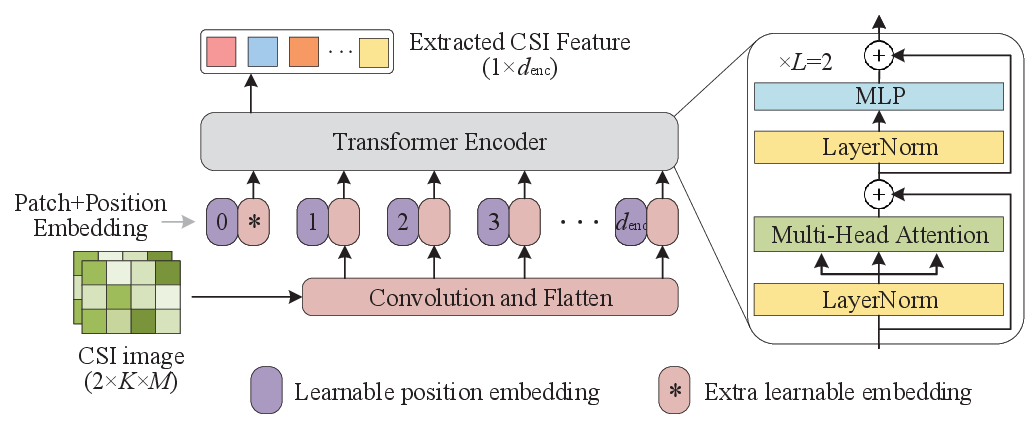}\\
\caption{The architecture of the adopted ViT model.}
\label{fig:vit_model}
\end{figure}

The CSI encoder module comprises three components: 1) a vision Transformer (ViT) model that extracts high-level features from CSI images, enabling the LLM to better interpret the CSI data; 2) a two-layer MLP model that maps the extracted feature vectors into embedding representations compatible with the LLM's textual input format; and 3) a positional embedding layer that incorporates temporal information into the CSI embeddings.
The adopted ViT model is illustrated in Fig. \ref{fig:vit_model}, which follows the architecture presented in \cite{dosovitskiy2020image}. Let $d_{\rm{enc}}$ denote the vector dimension of the Transformer encoder layers in the ViT model.
Given a CSI image $\mathbf{X}_t^{\rm{pre}} \in \mathbb{R}^{2\times K \times N}$, we first apply a convolutional layer to reshape and project $\mathbf{X}_t^{\rm{pre}}$ into a sequence of 2D patches, i.e.,
\begin{align}
\mathbf{X}_{t}^{\rm{patch}} = \text{Conv}(\mathbf{X}_t^{\rm{pre}}) \in \mathbb{R}^{d_{\rm{enc}} \times (K/P) \times (N/P)},
\end{align}
where $P$ represents the patch size, both the kernel size and stride of the convolutional layer are set to $(P, P)$, and the number of kernels is set as $d_{\rm{enc}}$.
This results in $M = \frac{K \times N}{P^2}$ patches, which defines the effective input sequence length for the Transformer encoder layers of the ViT model. These patches are then flattened into a sequence of $M$ patch embedding vectors, i.e.,
\begin{align}
[\mathbf{e}_{1}^{\rm{enc}}; \mathbf{e}_{2}^{\rm{enc}}; \cdots; \mathbf{e}_{M}^{\rm{enc}}] = \text{Flatten} (\mathbf{X}_{t}^{\rm{patch}}) \in \mathbb{R}^{M \times d_{\rm{enc}}},
\end{align}
where $\mathbf{e}_{i}^{\rm{enc}} \in \mathbb{R}^{1 \times d_{\rm{enc}}}$ ($\forall i = 1, 2, \cdots, M$) is the $i$-th patch embedding vector.

To extract a representative feature of the CSI image $\mathbf{X}_t^{\rm{pre}}$, we prepend a learnable embedding vector $\mathbf{e}_{\rm{feature}} \in \mathbb{R}^{1 \times d_{\rm{enc}}}$ to the sequence of patch embeddings.
The state of $\mathbf{e}_{\rm{feature}}$ at the output of the Transformer encoder layers is used as the final representation of the CSI image.
Then, learnable positional embeddings $\mathbf{E}_{\rm{pos}} \in \mathbb{R}^{(M+1) \times d_{\rm{enc}}}$ are then added to these embedding vectors to retain spatial information, i.e.,
\begin{align}
\mathbf{X}_t^{\rm{enc}}= [\mathbf{e}_{\rm{feature}}; \mathbf{e}_{1}^{\rm{enc}}; \mathbf{e}_{2}^{\rm{enc}}; \cdots; \mathbf{e}_{M}^{\rm{enc}}] + \mathbf{E}_{\rm{pos}}.
\end{align}
The resulting sequence of embedding vectors $\mathbf{X}_t^{\rm{enc}}$ is fed into the Transformer encoder layers.
The ViT model includes $L$ Transformer encoder layers, as shown in Fig. \ref{fig:vit_model}. Each Transformer encoder layer consists of a multi-head self-attention block and an MLP block, where layer normalization is applied before each block and residual connections are added after each block.
The last Transformer encoder layer outputs the encoded feature vectors for all input embeddings, i.e.,
\begin{align}
\mathbf{X}_t^{\rm{trans}}= \text{TransformerEncoder} (\mathbf{X}_t^{\rm{enc}}) \in \mathbb{R}^{(M+1) \times d_{\rm{enc}}},
\end{align}
which preserves the same shape as the input sequence.
We extract the first output vector as the feature vector of the CSI image $\mathbf{X}_t^{\rm{pre}}$, i.e.,
\begin{align}
\mathbf{x}_t^{\rm{feature}}= \mathbf{X}_t^{\rm{trans}}[1, :] \in \mathbb{R}^{1 \times d_{\rm{enc}}}.
\end{align}

After extracting high-level features from all CSI images, we obtain a sequence of feature vectors representing the CSI images, i.e., $[\mathbf{x}_t^{\rm{feature}}: t \in \{t_0-T_{\rm{P}}+1, \cdots, t_0\}]$. However, the dimensionality of these feature vectors ($d_{\rm{enc}}$) may not match the embedding dimension expected by the LLM ($d_{\rm{LLM}}$), i.e., $d_{\rm{enc}} \ne d_{\rm{LLM}}$. To address this problem, a two-layer MLP network is employed to transform each CSI feature vectors into CSI token vectors, aligning their dimensionality with that of the LLM's text embeddings. Specifically, $\mathbf{x}_t^{\rm{feature}}$ is mapped as
\begin{align}
\mathbf{x}_t^{\rm{token}} = \text{MLP}(\mathbf{x}_t^{\rm{feature}}) \in \mathbb{R}^{1 \times d_{\rm{LLM}}}.
\end{align}
Until now, we obtained a sequence of CSI token vectors for the CSI matrices of past $T_{\rm{P}}$ time slots, i.e., $[\mathbf{x}_t^{\rm{token}}: t \in \{t_0-T_{\rm{P}}+1, \cdots, t_0\}]$.
Since these CSI token vectors are computed without considering any temporal information, we further apply position embeddings as the indicator of temporal information to these CSI token vectors. In this work, we employ sinusoidal positional encoding \cite{takase2019positional} to encode the temporal position of each CSI token vector.
Specifically, for each CSI token vector $\mathbf{x}_t^{\rm{token}}$, we generate a positional embedding vector $\mathbf{e}_t^{\rm{PE}} = [e_{t, 1}^{\rm{PE}}, e_{t, 2}^{\rm{PE}}, \ldots, e_{t, d_{\rm{LLM}}}^{\rm{PE}}]$, where each element $e_{t,i}^{\rm{PE}}$ ($i \in \{1, 2, \cdots, d_{\rm{LLM}}\}$) is calculated as follows:
\begin{align}
e_{t,i}^{\rm{PE}} = \left\{ \begin{array}{*{20}{c}}
\sin \left( \frac{t}{10000^{i/d_{\rm{LLM}}}} \right)& \text{if}~ i~\text{mod}~2 = 0,\\
\cos \left( \frac{t}{10000^{(i - 1)/{d_{\rm{LLM}}}} }\right) & \text{Otherwise}.
\end{array} \right.
\end{align}
Then, we add the position embedding vectors into the CSI token vectors and obtain the token embedding vectors as
\begin{align}
\mathbf{x}_t^{\rm{emb}} = \mathbf{x}_t^{\rm{token}} + \mathbf{e}_t^{\rm{PE}}  \in \mathbb{R}^{1 \times d_{\rm{LLM}}}.
\end{align}

\begin{figure}[t]
\centering
\includegraphics[width=0.42\textwidth]{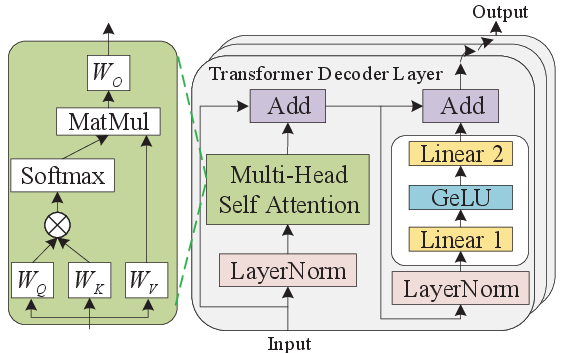}\\
\caption{The architecture of OPT-350m.}
\label{fig:OPT_architecture}
\end{figure}

3) \textbf{LLM backbone module}:
By applying the preprocessing and CSI encoder modules to the CSI data from the past $T_{\rm{P}}$ time slots, we obtain a sequence of $T_{\rm{P}}$ CSI token embedding vectors, i.e., $\mathbf{X}^{\rm{emb}} = [\mathbf{x}_{t_0 - T_{\rm{P}} +1}^{\rm{emb}}, \cdots, \mathbf{x}_{t_0}^{\rm{emb}}]$. This sequence is then fed into the LLM to generate the predicted CSI for the future $T_F$ time slots. The LLM used in this work is OPT-350M \cite{zhang2022opt}, which consists of stack Transformer decoder layers, as shown in Fig. \ref{fig:OPT_architecture}. Notably, the proposed CPLLM supports flexible substitution of the LLM backbone with other models, such as LLaMA \cite{touvron2023llama}. The selection of the LLM requires striking a balance between computational costs and prediction accuracy. Given the configuration of the LLM backbone, we input the sequence of CSI embedding vectors into the LLM and obtain the output as follows:
\begin{align}
\mathbf{X}_{\rm{LLM}}^{\rm{out}} = \text{LLM}( \mathbf{X}^{\rm{emb}}) \in \mathbb{R}^{T_{\rm{P}} \times d_{\rm{LLM}}}.
\end{align}

4) \textbf{CSI decoder module}:
Based on the output of the LLM, i.e., $\mathbf{X}_{\rm{LLM}}^{\rm{out}} $, we generate the predicted CSI for future $T_{\rm{F}}$ time slots.
Most popular LLMs (particularly decoder-only models such as OPT and LLaMA) employ an auto-regressive decoding strategy that generates tokens sequentially, which may lead to increased computational cost and inference latency in channel prediction. To address this issue, we design a CSI decoder module that generates the predicted CSI for future $T_{\rm{F}}$ time slots in parallel from $\mathbf{X}_{\rm{LLM}}^{\rm{out}}$, where the CSI matrix in each slot contains $2 \times K \times N$ elements. The CSI decoder module consists of an MLP network followed by a denormalization operator. The MLP network includes two fully-connected layers, each followed by a reshape operation, to generate $T_{\rm{F}} \times 2 \times K \times N$ values corresponding to the CSI matrices for the future $T_{\rm{F}}$ slots, i.e.,
\begin{align}
\mathbf{X}^{\rm{dec}} = \text{MLP}(\mathbf{X}_{\rm{LLM}}^{\rm{out}}) \in \mathbb{R}^{T_{\rm{F}} \times 2 \times K \times N},
\end{align}
where the first dimension corresponds to $T_{\rm{F}}$ CSI images, and the first and second rows of the second dimension represent the real and imaginary parts of each CSI image, respectively.
Then, we de-normalize $\mathbf{X}^{\rm{dec}}$ to obtain the predicted CSI images of future $T_{\rm{F}}$ slots, i.e.,
\begin{align}
\widehat{\mathbf{H}}^{\rm{real}} = \sigma_b \mathbf{X}^{\rm{dec}} + \mu_b.
\end{align}
Finally, the predicted CSI in future $T_{\rm{F}}$ slots is given by
\begin{align}\label{eq:CSI_real_to_imag}
\widehat{\mathbf{H}} = \widehat{\mathbf{H}}^{\rm{real}}[:,1,:,:] + j \widehat{\mathbf{H}}^{\rm{real}}[:,2,:,:] \in \mathbb{R}^{T_{\rm{F}} \times K \times N},
\end{align}
where $j=\sqrt{-1}$ is the imaginary unit, $\widehat{\mathbf{H}}_t = \widehat{\mathbf{H}}[t, :, :]$ is the predicted CSI in time slot $t$ ($t \in \{t_0 + 1, \cdots, t_0 + T_{\rm{F}} \}$).

For clarity, we summarize the key steps of the CPLLM channel prediction inference process in Algorithm \ref{alg:inference_CP} to illustrate the data flow of CSI prediction.

\begin{algorithm}[t] \small
\caption{Inference of CPLLM}
\label{alg:inference_CP}
\begin{algorithmic}[1]
\State \textbf{Inputs:} Historical CSI data $[\mathbf{H}_{t_0-T_{\rm{P}}+1}, \cdots, \mathbf{H}_{t_0}]$.
\For{$\mathbf{H}_t \in [\mathbf{H}_{t_0-T_{\rm{P}}+1}, \cdots, \mathbf{H}_{t_0}]$}
    \State Preprocessing $\mathbf{H}_t$ to obtain the normalized CSI image $\mathbf{X}_{t}^{\rm{pre}}$.
    \State Pass $\mathbf{X}_{t}^{\rm{pre}}$ to the CSI encoder to obtain a token vector $\mathbf{x}_{t}^{\rm{token}}$.
\EndFor
\State Add the position embedding vectors into CSI token vectors and obtain $T_{\rm{P}}$ token embedding vectors, i.e., $\mathbf{X}^{\rm{emb}}$.
\State Feed $\mathbf{X}^{\rm{emb}}$ to the LLM backbone and obtain output as $\mathbf{X}_{\rm{LLM}}^{\rm{out}}$.
\State Pass $\mathbf{X}_{\rm{LLM}}^{\rm{out}}$ to the CSI decoder and obtain the predicted CSI images for future $T_{\rm{F}}$ slots, i.e., $\widehat{\mathbf{H}}^{\rm{real}}$.
\State Convert $\widehat{\mathbf{H}}^{\rm{real}}$ into complex-value CSI according to \eqref{eq:CSI_real_to_imag}.
\end{algorithmic}
\end{algorithm}

\subsection{Fine-tuning of CPLLM Model}
In this subsection, we fine-tune the proposed CPLLM model, which comprises parameters from the CSI encoder, the pretrained LLM backbone, and the CSI decoder. Among these, the LLM backbone accounts for the majority of the model's parameters, while the contributions from the CSI encoder and decoder are comparatively negligible. Since LLMs typically contain a vast number of parameters, training a channel prediction-specific LLM from scratch to match the reasoning and generalization capabilities of well-established LLMs such as OPT \cite{zhang2022opt} and LLaMA \cite{touvron2023llama} is costly and challenging. Fortunately, numerous existing works, e.g., \cite{NEURIPS2023_6dcf277e}, have demonstrated that fine-tuning pretrained LLMs on task-specific datasets enables them to achieve impressive few-shot and zero-shot generalization capabilities on downstream tasks. Motivated by this, we adopt a parameter-efficient fine-tuning approach, i.e., Low-Rank Adaptation (LoRA) \cite{hu2022lora}, to adapt the pretrained LLM for the channel prediction task.

Specifically, we focus on fine-tuning the weights of the query, key, and value projection layers in each Transformer decoder layer of the LLM, as shown in Fig. \ref{fig:OPT_architecture}. For each projection layer with parameter matrix $\mathbf{W}$, LoRA injects trainable low-rank decomposition matrices, i.e., $\mathbf{A}$ and $\mathbf{B}$, into the layer. As a result, the original weight $\mathbf{W}$ is modified as:
\begin{align}
\mathbf{W} \leftarrow \mathbf{W} + \frac{\alpha }{r}\mathbf{B} \mathbf{A},
\end{align}
where $\mathbf{W} \in \mathbb{R}^{d_{\rm{LLM}} \times d_{\rm{LLM}}}$, $\mathbf{A} \in \mathbb{R}^{d_{\rm{LLM}} \times r}$, $\mathbf{B} \in \mathbb{R}^{r \times d_{\rm{LLM}}}$, $r \ll d_{\rm{LLM}}$ is the rank of LoRA adapter, and $\alpha$ is the scaling factor to adjust the importance of LoRA adapter. During the fine-tuning process, the weight matrix $\mathbf{W}$ in the pretrained LLM remains frozen and is not updated, while the LoRA adapter matrices $\mathbf{A}$ and $\mathbf{B}$ are trainable. In general, the weights of $\mathbf{A}$ are initialized following a Gaussian distribution, while $\mathbf{B}$ is initialized with all zeros.
Since $r \ll d_{\rm{LLM}}$, the number of trainable parameters introduced by $\mathbf{A}$ and $\mathbf{B}$ (i.e., $2rd_{\rm{LLM}}$) is significantly smaller than that of $\mathbf{W}$ (i.e., $d_{\rm{LLM}}^2$). Consequently, LoRA makes the fine-tuning process both efficient and affordable. Moreover, LoRA helps prevent catastrophic forgetting of the LLM's original knowledge, enabling us to leverage the universal reasoning and generalization capabilities of pretrained LLMs for channel prediction.

With the assistance of LoRA, we introduce LoRA adapters into the query, key, and value projection layers of each Transformer decoder layer in the LLM backbone. We then fine-tune CPLLM to learn the spatial and temporal correlation of downlink CSI, enabling accurate prediction of future CSI over $T_{\rm{F}}$ time slots given the CSI from the past $T_{\rm{P}}$ slots as input. During the fine-tuning process, the LLM backbone remains frozen, while the parameters of the CSI encoder, the LoRA adapters, and the CSI decoder are updated. Specifically, given a fine-tuning dataset $\{(\bm{\mathcal{H}}_i^{\rm{P}}, \bm{\mathcal{H}}_i^{\rm{F}}\})_{i=1}^M$ with $M$ data samples, where each sample $(\bm{\mathcal{H}}_i^{\rm{P}}, \bm{\mathcal{H}}_i^{\rm{F}}\})$ comprises input data $\bm{\mathcal{H}}_i^{\rm{P}} \in \mathbb{C}^{T_{\rm{P}} \times K \times N}$ that contains the CSI over the past $T_{\rm{P}}$ time slots, and the corresponding label $\bm{\mathcal{H}}_i^{\rm{F}} \in \mathbb{C}^{T_{\rm{F}} \times K \times N}$ that contains the CSI for the future $T_{\rm{F}}$ time slots.
In each fine-tuning epoch, we randomly sample a batch of data samples $\mathcal{B} = \{ (\bm{\mathcal{H}}_b^{\rm{P}}, \bm{\mathcal{H}}_b^{\rm{F}}) \}_{b=1}^B$ with batch size $B$ from the fine-tuning dataset.
Then, the batch of input data $\{\bm{\mathcal{H}}_b^{\rm{P}}\}_{b=1}^B$ is fed into CPLLM to generate predicted CSI, denoted as $\{\widehat{\bm{\mathcal{H}}}_b^{\rm{F}}\}_{b=1}^B$.
Hence, the model is fine-tuned by minimizing the loss function as follows:
\begin{align}\label{eq:loss_cp}
\mathcal{L}_{\rm{CP}} = \frac{1}{B}\sum\nolimits_{b = 1}^B \frac{\big\| \bm{\mathcal{H}}_b^{\rm{F}}  - \widehat{\bm{\mathcal{H}}}_b^{\rm{F}} \big\|_F^2} {\big\| \bm{\mathcal{H}}_b^{\rm{F}} \big\|_F^2}.
\end{align}
The loss function quantifies the difference between the predicted CSI $\{\widehat{\bm{\mathcal{H}}}_b^{\rm{F}}\}_{b=1}^B$ and the ground truth $\{\bm{\mathcal{H}}_b^{\rm{F}}\}_{b=1}^B$. Finally, the trainable parameters in the model are updated using gradient descent based on the loss.
For clarity, the detailed steps for fine-tuning the proposed LLM-based channel prediction model are summarized in Algorithm \ref{alg:finetuning_CP}.

\begin{algorithm}[t] \small
\caption{Fine-tuning of CPLLM}
\label{alg:finetuning_CP}
\begin{algorithmic}[1]
\State \textbf{Inputs:} The fine-tuning dataset $\{(\bm{\mathcal{H}}_i^{\rm{P}}, \bm{\mathcal{H}}_i^{\rm{F}}\})_{i=1}^M$, LoRA adapter rank $r$, LoRA scaling factor $\alpha$, batch size $B$.
\State Create LoRA adapters for the query, key, and value projection layers in each Transformer decoder layer of the LLM backbone.
\State Freeze the LLM backbone and enable the parameters of the CSI encoder, LoRA adapter, and the CSI decoder to be trainable.
\Repeat
    \State \parbox[t]{\dimexpr\linewidth-\algorithmicindent}{Randomly sample a batch of $B$ data samples from the fine-tuning dataset, i.e., $\mathcal{B} = \{ (\bm{\mathcal{H}}_b^{\rm{P}}, \bm{\mathcal{H}}_b^{\rm{F}}) \}_{b=1}^B$.}
    \State \parbox[t]{\dimexpr\linewidth-\algorithmicindent}{Forward pass $\{ \bm{\mathcal{H}}_b^{\rm{P}} \}_{b=1}^B$ through the channel prediction model and generate predicted CSI as $\{\widehat{\bm{\mathcal{H}}}_b^{\rm{F}}\}_{b=1}^B$.}
    \State Compute the fine-tuning loss according to \eqref{eq:loss_cp}.
    \State \parbox[t]{\dimexpr\linewidth-\algorithmicindent}{Perform gradient descent based on the computed loss to update parameters of the CSI encoder, LoRA adapters, and the CSI decoder in CPLLM.}
\Until{the CPLLM model converged.}
\end{algorithmic}
\end{algorithm}

\section{LLM-Based Predictive Beamforming}\label{sec:beamforming}
By leveraging the proposed CPLLM described in Section \ref{sec:channel_prediction} to perform channel prediction, the CSI for the future $T_{\rm{F}}$ slots can be obtained. For each future slot, the beamforming optimization problem \eqref{prob:P} can then be formulated, and existing methods, e.g., \cite{4599181, 4712693, 9000850, 10550141}, can be employed to derive the corresponding beamforming strategy.
However, the sequential paradigm of first performing channel prediction followed by beamforming increases the signal processing overhead in LEO satellite communication systems. In addition, due to the discrepancy between the predicted CSI and the perfect CSI, the sequential paradigm may cause error propagation, further degrading the performance of the obtained beamforming strategy.
To this end, this section further extends the proposed CPLLM to BFLLM, which directly predicts the transmit beamforming strategy based on historical downlink CSI, aiming to reduce signal processing overhead and improve downlink transmission performance for LEO satellite communications.

\subsection{Problem Formulation for Predictive Beamforming}
This section aims to leverage the historical downlink CSI of devices over the past $T_{\rm{P}}$ time slots, i.e., $[\mathbf{H}_{t_0-T_{\rm{P}}+1}, \cdots, \mathbf{H}_{t_0}]$, to directly predict the beamforming strategies for the upcoming $T_{\rm{F}}$ time slots. To this end, we formulate the predictive beamforming problem as follows:
\begin{align}
\max_{\bm{\Theta}}&~ \frac{1}{T_{\rm{F}}} \sum\limits_{t = 1}^{T_{\rm{F}}} \sum\limits_{k = 1}^K \log_2\left( 1 + \Gamma_{k}(t_0 + t, f) \right) \label{prob:predictive_beamforming} \\
\text{s.~t.~} & \sum\nolimits_{k = 1}^K \left\| \widehat{\mathbf{w}}_{k,t_0 + t} \right\|^2  \le P_T, \forall t \in \{1, \cdots, T_{\rm{F}}\}, \tag{\theequation a} \\
& \big[ \widehat{\mathbf{W}}_{t_0 + 1}, \cdots , \widehat{\mathbf{W}}_{t_0 + T_{\rm{F}}} \big] = f_{\bm{\Theta}} (\left[ \mathbf{H}_{t_0-T_{\rm{P}}+1}, \cdots, \mathbf{H}_{t_0} \right]), \tag{\theequation b}
\end{align}
where $\widehat{\mathbf{W}}_t = [\widehat{\mathbf{w}}_{1,t}, \widehat{\mathbf{w}}_{2,t}, \cdots, \widehat{\mathbf{w}}_{K,t}]^T \in \mathbb{C}^{K \times N}$ is the predicted beamforming strategy in time slot $t$, $f_{\bm{\Theta}}(\cdot)$ is the predictive beamforming model,  $\bm{\Theta}$ is the set of model parameters.

\subsection{LLM-Based Predictive Beamforming Framework}
In this subsection, we design an LLM-based predictive beamforming model, referred to as BFLLM, to address problem \eqref{prob:predictive_beamforming}. It follows the same architecture as the CPLLM model (described in Section \ref{subsec:CP_LLM}), sharing identical preprocessing, CSI encoder, and LLM backbone components. The main difference lies in the operations of the decoder module.
Given the historical CSI data over the past $T_{\rm{P}}$ time slots, i.e., $\bm{\mathcal{H}}_{\rm{P}} = [\mathbf{H}_{t_0-T_{\rm{P}}+1}, \cdots, \mathbf{H}_{t_0}]$, we feed them into the BFLLM model by sequentially passing them through the preprocessing module, the CSI encoder, and the LLM backbone, ultimately obtaining the output $\mathbf{Y}_{\rm{LLM}}^{\rm{out}} \in  \mathbb{R}^{T_{\rm{P}} \times d_{\rm{LLM}}}$, i.e.,
\begin{align}
\mathbf{Y}_{\rm{LLM}}^{\rm{out}} = \text{LLM}(\text{CSIEncoder}(\text{Preprocssing}(\bm{\mathcal{H}}_{\rm{P}}))) .
\end{align}
Then, $\mathbf{Y}_{\rm{LLM}}^{\rm{out}}$ is fed into the beamforming decoder module to generate the predicted beamforming strategy for the future $T_{\rm{F}}$ time slots.
The beamforming decoder module consists of an MLP network and a scaling operator. The MLP network shares the same architecture as the one in the CSI decoder module described in Section \ref{subsec:CP_LLM}, consisting of two fully connected layers, each followed by a reshape operation.
Specifically, the MLP network maps $\mathbf{Y}_{\rm{LLM}}^{\rm{out}}$ into $T_{\rm{F}} \times 2 \times K \times N$ values that correspond to the beamforming matrices for future $T_{\rm{F}}$ time slots, i.e.,
\begin{align}
\mathbf{Y}^{\rm{dec}} = \text{MLP}(\mathbf{Y}_{\rm{LLM}}^{\rm{out}}) \in \mathbb{R}^{T_{\rm{F}} \times 2 \times K \times N},
\end{align}
where $\mathbf{Y}^{\rm{dec}}[t,:,:,:]$ corresponds to the predicted beamforming matrix in the $t$-th future slot.
Then, we scale each $\mathbf{Y}^{\rm{dec}}[t,:,:,:]$ ($t \in \{1, 2, \cdots, T_{\rm{F}}\}$) to meet the power constraint and obtain the predicted beamforming strategy in each slot as follows:
\begin{align}
\widehat{\mathbf{W}}_{t_0 + t}^{\rm{real}} = \frac{\mathbf{Y}^{\rm{dec}}[t,:,:,:]} {\left\| \mathbf{Y}^{\rm{dec}}[t,:,:,:] \right\|^2} P_T, \forall t \in \{1, 2, \cdots, T_{\rm{F}}\}.
\end{align}
Finally, the predicted beamforming matrix in future slot $t$ is $\widehat{\mathbf{W}}_{t_0 + t} = \widehat{\mathbf{W}}_{t_0 + t}^{\rm{real}}[1,:,:] + j\widehat{\mathbf{W}}_{t_0 + t}^{\rm{real}}[2,:,:]$.

\subsection{Fine-tuning of BFLLM Model}
The fine-tuning process of the proposed LLM-based BFLLM model follows the same procedure as that of the CPLLM model. Specifically, LoRA adapters are inserted into the query, key, and value projection layers of each Transformer decoder layer in the pretrained LLM backbone. During fine-tuning, the LLM backbone is kept frozen, while only the parameters of the CSI encoder, LoRA adapters, and beamforming decoder are updated. The fine-tuning process of the BFLLM model leverages the same dataset used in the channel prediction task, i.e., $\{(\bm{\mathcal{H}}_i^{\rm{P}}, \bm{\mathcal{H}}_i^{\rm{F}}\})_{i=1}^M$ with $M$ data samples. In each fine-tuning epoch, a batch of data samples $\mathcal{B} = \{ (\bm{\mathcal{H}}_b^{\rm{P}}, \bm{\mathcal{H}}_b^{\rm{F}}) \}_{b=1}^B$ with batch size $B$ is randomly sampled from the dataset. Next, the batch of historical CSI data $\{\bm{\mathcal{H}}_b^{\rm{P}}\}_{b=1}^B$ is then fed into the BFLLM model to generate the predicted beamforming matrices for the future $T_{\rm{F}}$ time slots, denoted as $\{\widehat{\mathbf{W}}_b\}_{b=1}^B$, where $\widehat{\mathbf{W}}_b \in \mathbb{C}^{T_{\rm{F}} \times K \times N}$ corresponds to the predictive results for the $b$-th input data sample. Then, we calculate the beamforming loss as follows:
\begin{align}
\mathcal{L}_{\rm{BF}} =  - \frac{1}{B T_{\rm{F}}} \sum\limits_{b = 1}^B \sum\limits_{t = 1}^{T_{\rm{F}}} \sum\limits_{k = 1}^K \log_2\left( 1 + \Gamma_{b, t, k} \right),
\end{align}
where
\begin{align}
\Gamma_{b,t,k} = \frac{\left| \bm{\mathcal{H}}_b^{\rm{F}}[t,k,:]^H \widehat{\mathbf{W}}_b[t,k,:] \right|^2}  {\sum\nolimits_{j = 1,j \ne k}^K \left| \bm{\mathcal{H}}_b^{\rm{F}}[t,k,:]^H \widehat{\mathbf{W}}_b[t,j,:] \right|^2 + \sigma^2 }.
\end{align}
Finally, the BFLLM model is updated using gradient descent based on the computed loss.

\section{Simulation Results}\label{sec:simulations}
This section evaluates the effectiveness of the proposed CPLLM and BFLLM for channel prediction and predictive beamforming, respectively. Unless otherwise specified, the default experimental parameters for the LEO satellite system are summarized in Table \ref{tab:simu_setting}, which are set up based on the 3GPP TR 38.821 specifications\footnote{\url{https://www.3gpp.org/ftp/Specs/archive/38_series/38.821/}}. For fine-tuning and testing the proposed CPLLM and BFLLM models, we first generate a fine-tuning dataset comprising 9,000 samples according to the channel model presented in Section \ref{sec:system_model}, with device speed uniformly selected from 10 km/h to 100 km/h.
Subsequently, we generate a test dataset of 1,000 samples using 10 discrete device velocities ranging from 10 km/h to 100 km/h in 10 km/h increments. In both the fine-tuning and test datasets, each data sample consists of the CSI of all devices over $T_{\rm{P}}+T_{\rm{F}}=20$ time slots.
In addition, following prior works in \cite{10582829, 10021887}, we add Gaussian white noise in the historical and future CSI in each training data sample, as well as historical CSI in each test data sample, to simulate channel estimation errors in acquiring these CSI data.
Specifically, each CSI sample in the training dataset is corrupted with Gaussian white noise, with the SNR uniformly sampled between 5 and 20 dB, while the historical CSI in the test dataset is corrupted with Gaussian white noise at an SNR of 15 dB by default, unless otherwise specified.

For both the CPLLM model and BFLLM model, the number of Transformer encoder layers in the CSI encoder module is set to 2, the hidden dimension of the Transformer encoder layer is set to $d_{\rm{enc}}=512$, and OPT-350M is adopted as their LLM backbones. During the fine-tuning process of the proposed  CPLLM model and BFLLM model, the LoRA rank is set to $r=8$, the LoRA scaling factor is set to $\alpha = 32$, the batch size to 64, and the AdamW optimizer is employed with a learning rate of 0.0001. Both models are trained for 300 epochs on the training dataset.
\begin{table}[ht]\small
\caption{Parameter Settings for LEO Satellite Communication System}
\centering
\label{tab:simu_setting}
\begin{tabular}{p{5.5cm}|p{1.3cm}}
\hline
Parameter & Value \\
\hline
Carrier frequency $f_c$   & 5 GHz\\
Attitude of LEO satellite & 600 km \\
Number of scatter paths $L_k$ & 6  \\
LEO satellite velocity & 7.5km/s \\
Rician factor $\kappa_k$ & 10 dB \\
Number of antennas $N$ & 16 \\
Number of devices $K$ & 10 \\
Number of historical slots $T_{\rm{P}}$ & 16 \\
Number of predicting slots $T_{\rm{F}}$ & 4 \\
Time interval between slots & 0.5 ms \\
Maximum delay spread $\max \{\Delta \tau_{k,l}\}$ & 30 ns \\
Noise variance $\sigma^2$ & -10 dBw \\
Total transmit power of LEO satellite $P_T$ & 0 dBw \\
\hline
\end{tabular}
\end{table}

\subsection{Effectiveness of Channel Prediction}
In this subsection, we evaluate the effectiveness of the proposed channel prediction approach, i.e., CPLLM, by comparing it with the following benchmark methods:
\begin{itemize}
  \item RNN-based channel prediction (RNN-CP) \cite{8813020}: RNN is a well-established architecture commonly utilized for sequence modeling and widely applied to channel prediction tasks. In our simulations, the RNN model is configured with four layers.
  \item LSTM-based channel prediction (LSTM-CP) \cite{9439942}: LSTM introduces gating mechanisms to overcome the limitations of RNNs, enabling effective preservation of information over long sequences. In our simulations, we configure the LSTM model with four layers for channel prediction.
  \item GRU-based channel prediction (GRU-CP) \cite{10089512}: GRU is a variant of LSTM that simplifies its architecture while retaining comparable performance in capturing temporal dependencies. In our simulations, the GRU model is also configured with four layers for channel prediction.
  \item Transformer-based channel prediction (Transformer-CP)\cite{9832933}: This approach leverages a Transformer network to predict CSI in future time slots in parallel, effectively mitigating the problem of error propagation.
  \item LSTM combined with CNN for channel prediction (LSTM+CNN-CP) \cite{9921297}: This approach combines LSTM and CSNN to capture both the spatial correlations and temporal dependencies of CSI for channel prediction.
  \item LLM4CP \cite{10582829}: LLM4CP is a pioneering LLM-based channel prediction approach that leverages GPT-2 as its backbone and achieves high prediction accuracy.
\end{itemize}

Consistent with prior works on channel prediction, e.g., \cite{10582829, 8813020, 8801923}, we use the NMSE to measure the channel prediction performance.

\begin{figure}[t]
\centering
\includegraphics[width=0.4\textwidth]{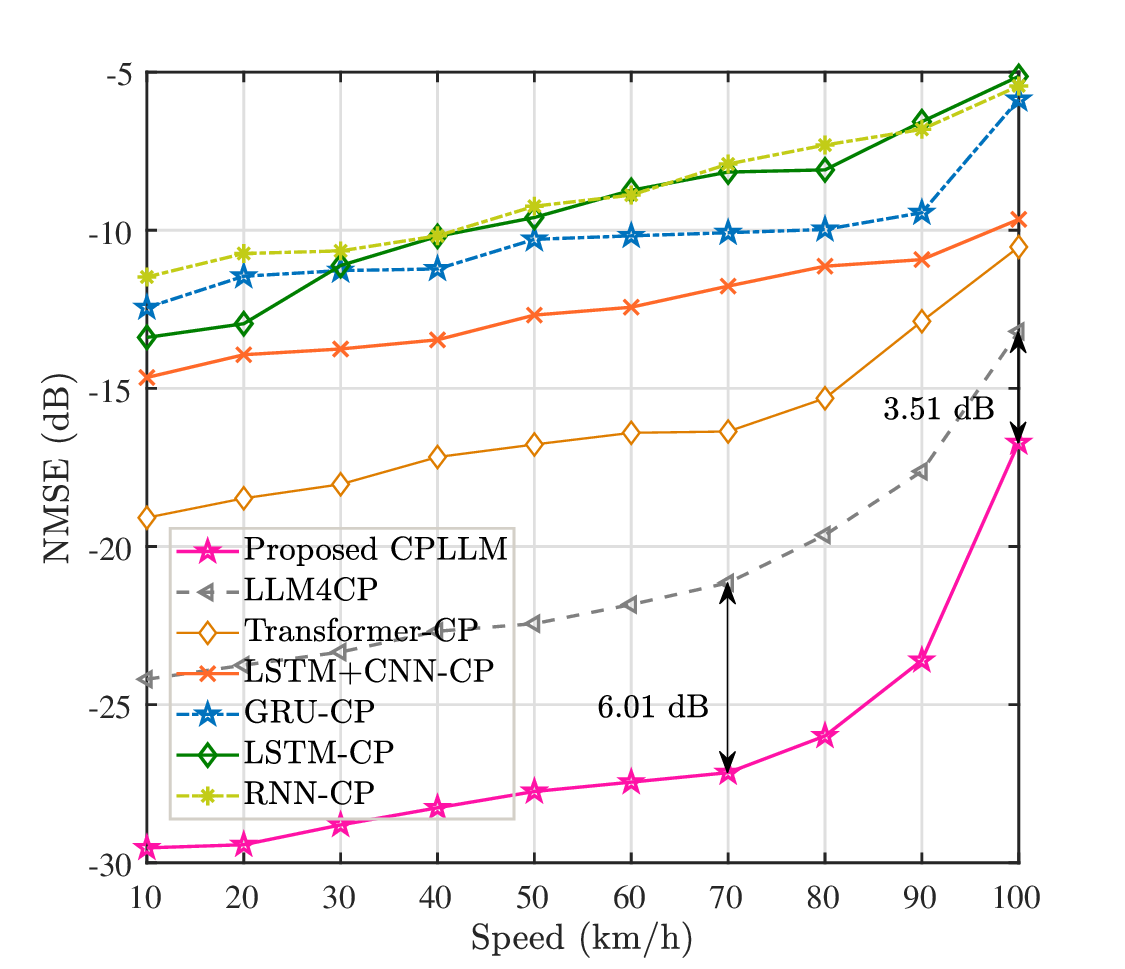}
\caption{Comparison of different channel prediction approaches across different device speeds.}
\label{fig:nmse_velocity}
\end{figure}

In Fig. \ref{fig:nmse_velocity}, we compare the NMSE performance of the proposed LLM-based channel prediction approach with the benchmarks under different device velocities. It is observed that the NMSE of all approaches increases with the growth of device speeds. This trend is attributed to higher velocities causing larger Doppler shifts and shorter channel coherence times, thereby increasing the complexity of accurate channel prediction.
In addition, the proposed LLM-based channel prediction method consistently outperforms all benchmark approaches across all device velocity settings. Compared to the LLM4CP method, the proposed approach achieves up to a 6 dB reduction in NMSE. This performance gain is primarily due to the proposed CPLLM's ability to simultaneously process CSI data from all transmit-receive antenna pairs across all devices, combined with parameter-efficient fine-tuning via LoRA. It enables the pretrained LLM to effectively capture and exploit both spatial and temporal channel correlations, leading to more accurate channel prediction.

\begin{figure}[t]
\centering
\includegraphics[width=0.4\textwidth]{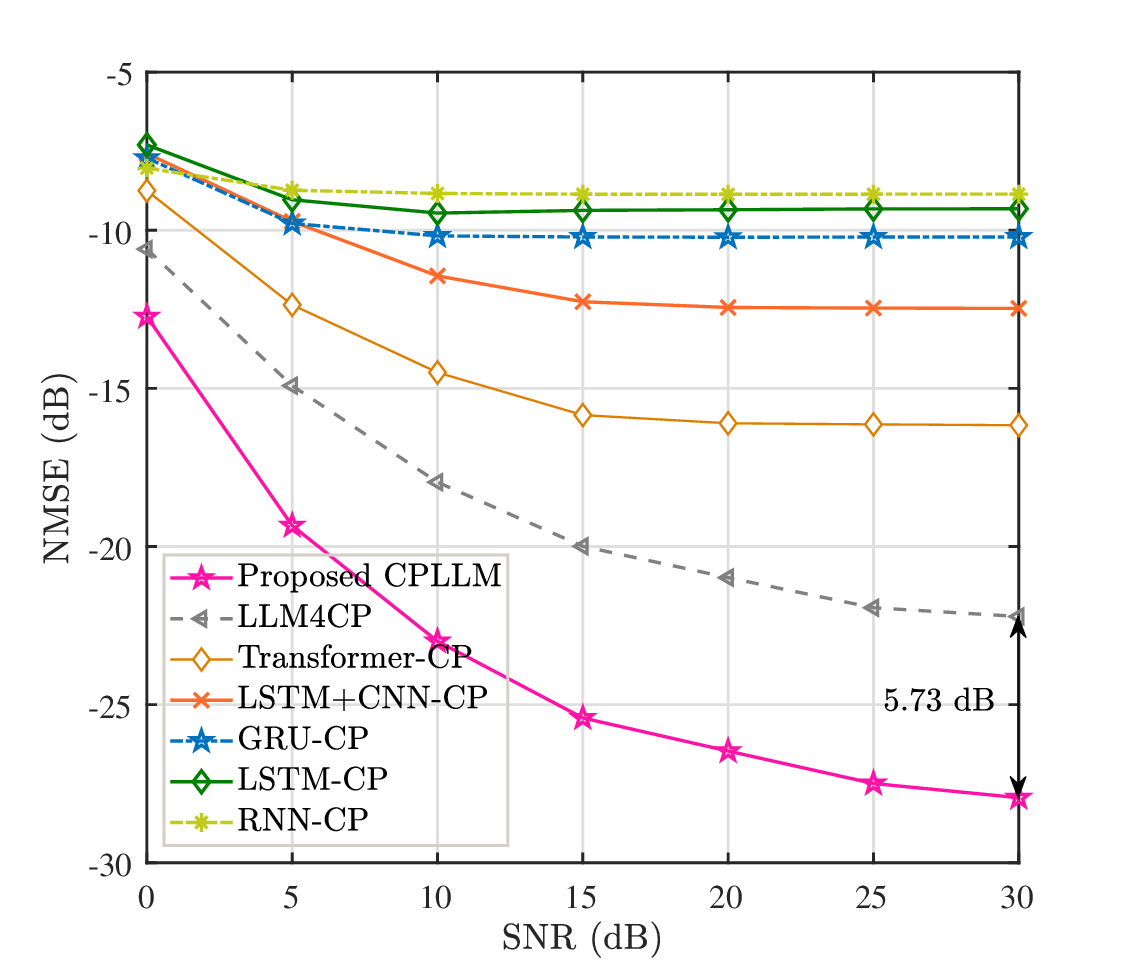}
\caption{Comparison of different channel prediction approaches across different SNRs.}
\label{fig:nmse_snr}
\end{figure}

Fig. \ref{fig:nmse_snr} evaluates the robustness of the proposed LLM-based channel prediction approach to the channel estimation errors in acquiring historical CSI data. To this end, we add Gaussian white noise into the historical CSI data in the test dataset, with SNR values ranging from 0 dB to 30 dB. Note that the results in Fig. \ref{fig:nmse_snr} are obtained by testing on the entire test dataset, which includes all 10 device velocity settings. It can be seen that the proposed CPLLM and the LLM4CP method significantly outperform the other benchmarks, which is attributed to the impressive reasoning and generalization capabilities of pretrained LLMs.
Moreover, the proposed approach outperforms all benchmarks across all SNR levels. Compared to the best benchmark, i.e., LLM4CP, it achieves up to 5.73 dB reduction in NMSE.

\begin{figure}[t]
\centering
\includegraphics[width=0.4\textwidth]{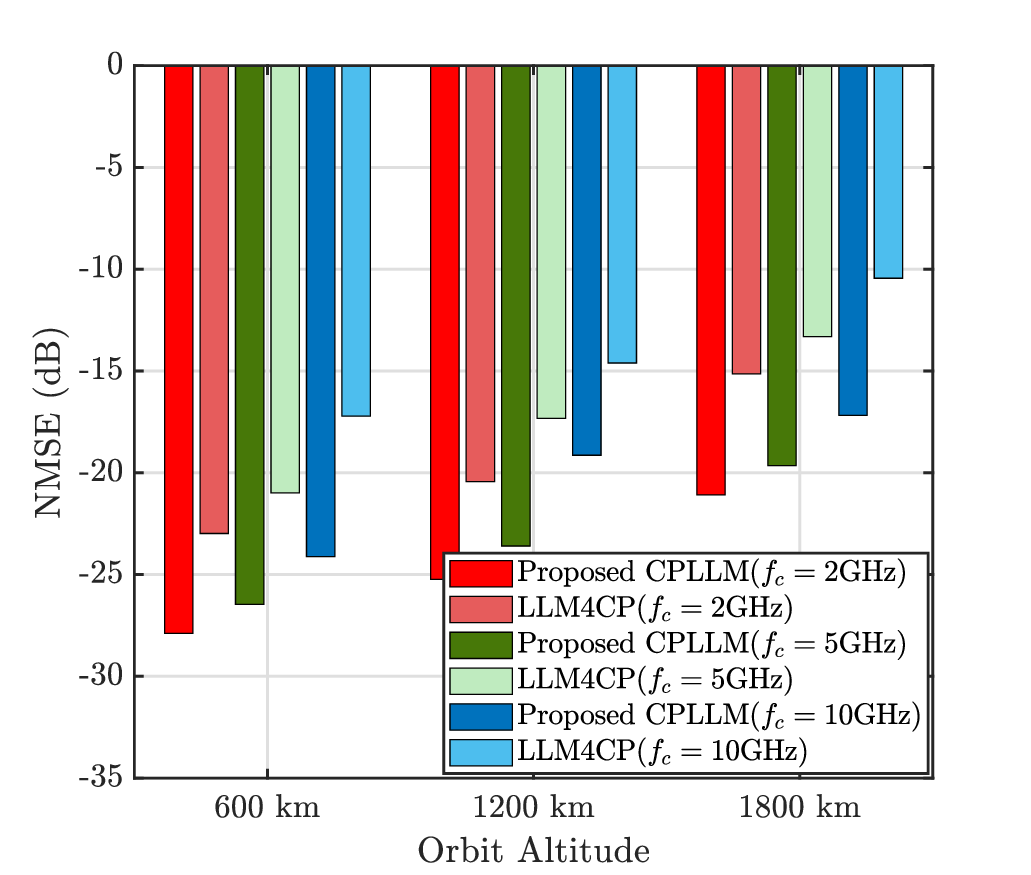}
\caption{Comparison of the proposed LLM-based channel prediction with LLM4CP under different carrier frequencies and orbit altitudes.}
\label{fig:nmse_altitude_frequency}
\end{figure}

In Fig. \ref{fig:nmse_altitude_frequency}, we compare the proposed LLM-based channel prediction approach with the state-of-the-art channel prediction approach, i.e., LLM4CP, under different LEO satellite altitudes and carrier frequencies. To this end, we generate a separate test dataset for each satellite altitude and carrier frequency setting. All other parameters for these datasets remain consistent with the default test dataset described at the beginning of this section, except for the satellite altitude and carrier frequency. Note that the model parameters for both the proposed approach and LLM4CP are the same as those used in the previous simulations. These parameters were obtained by fine-tuning on the fine-tuning dataset, without any retraining for specific satellite altitude or carrier frequency settings. It is observed that the NMSE increases with rising carrier frequency and orbit altitude. This is because higher carrier frequencies lead to greater Doppler shifts caused by both the devices and the LEO satellite, while increased orbit altitudes amplify the Doppler shift introduced by the satellite. Consequently, both factors reduce the channel coherence time, thereby making channel prediction more challenging. Additionally, both the proposed approach and LLM4CP demonstrate satisfactory NMSE performance across various carrier frequencies and orbit altitudes. It is worth noting that, except for the case of $f_c=5$ GHz and orbit altitude as 600 km, these environment settings were not seen during the fine-tuning process. Interestingly, under these unseen environments, their NMSE performance even surpasses the performance of other benchmark methods in Fig. \ref{fig:nmse_velocity}, which were evaluated without any training-test environment shift. This highlights the advantage of LLM-based channel prediction approaches, which benefit from the strong generalization capabilities inherited from pretrained LLMs. Moreover, consistent with the above results, the proposed approach showcases better NMSE performance than the LLM4CP across all settings.

\begin{figure}[t]
\centering
\includegraphics[width=0.4\textwidth]{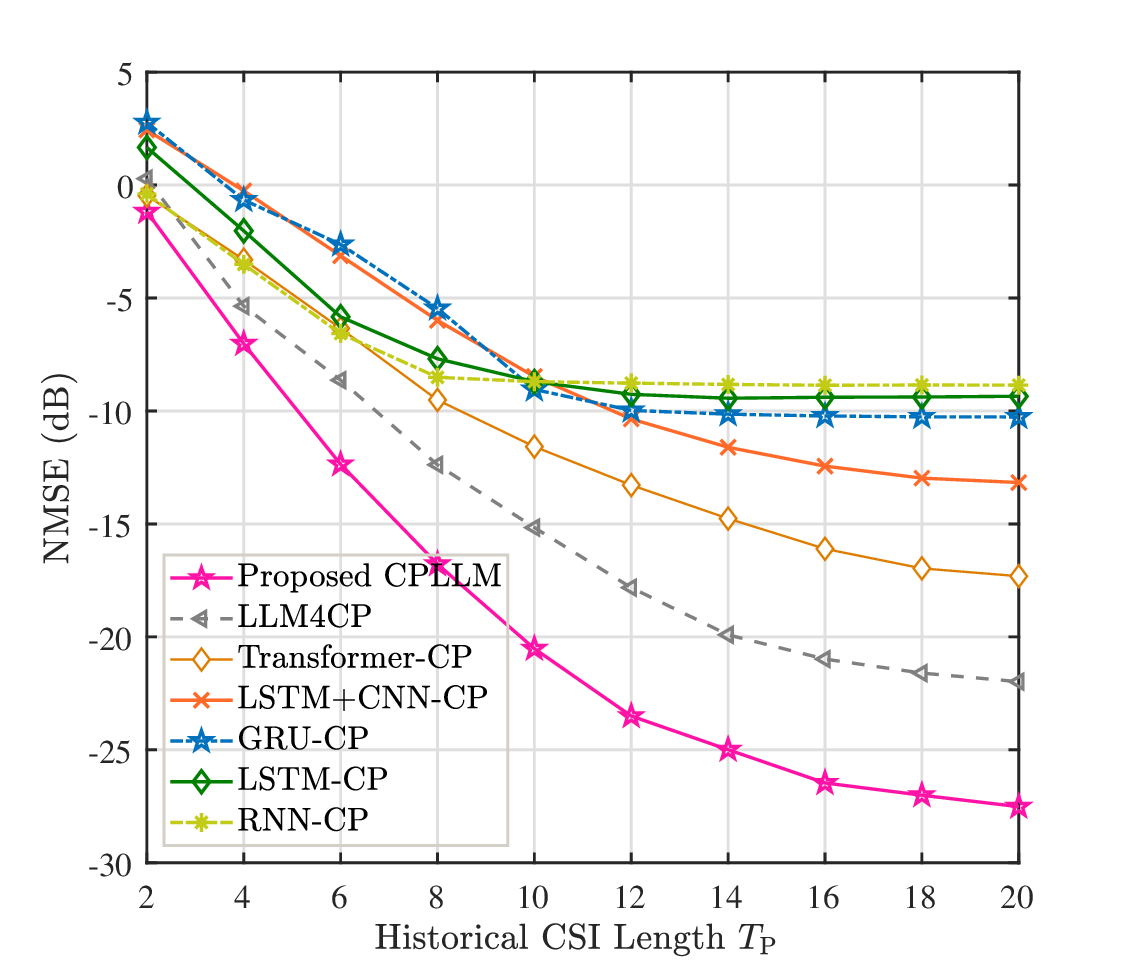}
\caption{Comparison of different channel prediction approaches under various historical CSI lengths.}
\label{fig:nmse_input_length}
\end{figure}

In Fig. \ref{fig:nmse_input_length}, we investigate the impact of the number of input historical CSI data on channel prediction performance. It is observed that the proposed approach consistently achieves the lowest NMSE compared to the benchmark methods. Additionally, the NMSE decreases as the number of historical CSI data increases, indicating that more historical information improves prediction accuracy. However, a longer historical CSI length would increase the computational costs for the channel prediction process. Therefore, selecting an appropriate input sequence length is crucial to balancing prediction performance and computational cost.

\subsection{Effectiveness of Predictive Beamforming}\label{subsec:simu_BF}
This section verifies the effectiveness of the proposed BFLLM by comparing it with the following benchmarks. Following previous works, e.g., \cite{10021887, 10138552, 10741218}, we adopt the sum rate to measure the predictive beamforming perfomance.
\begin{itemize}
  \item Sequential paradigm of LSTM-based channel prediction followed by MLP-based beamforming (LSTM+MLP-BF) \cite{10021887}: This approach first employs an LSTM model to predict future CSI from historical CSI. Then, an MLP model is trained to generate beamforming solution by using predicted CSI as its input.
  \item Integrated LSTM and CNN for predictive beamforming (LSTM+CNN-BF)\cite{10138552}: It combines CNN and LSTM to extract spatial and temporal features from historical CSI for predicting the beamforming strategy.
  \item Transformer-based predictive beamforming (Transformer-BF) \cite{10741218}: This approach adopts a Transformer model to directly predict the beamforming strategy by inputting historical CSI data.
  \item Sequential paradigm of the proposed CPLLM for channel prediction followed by WMMSE for beamforming (CPLLM+WMMSE-BF): This approach first utilizes the proposed CPLLM to forecast future CSI, followed by the use of WMMSE to compute the beamforming solutions based on the predicted CSI.
\end{itemize}

\begin{figure}[t]
\centering
\includegraphics[width=0.4\textwidth]{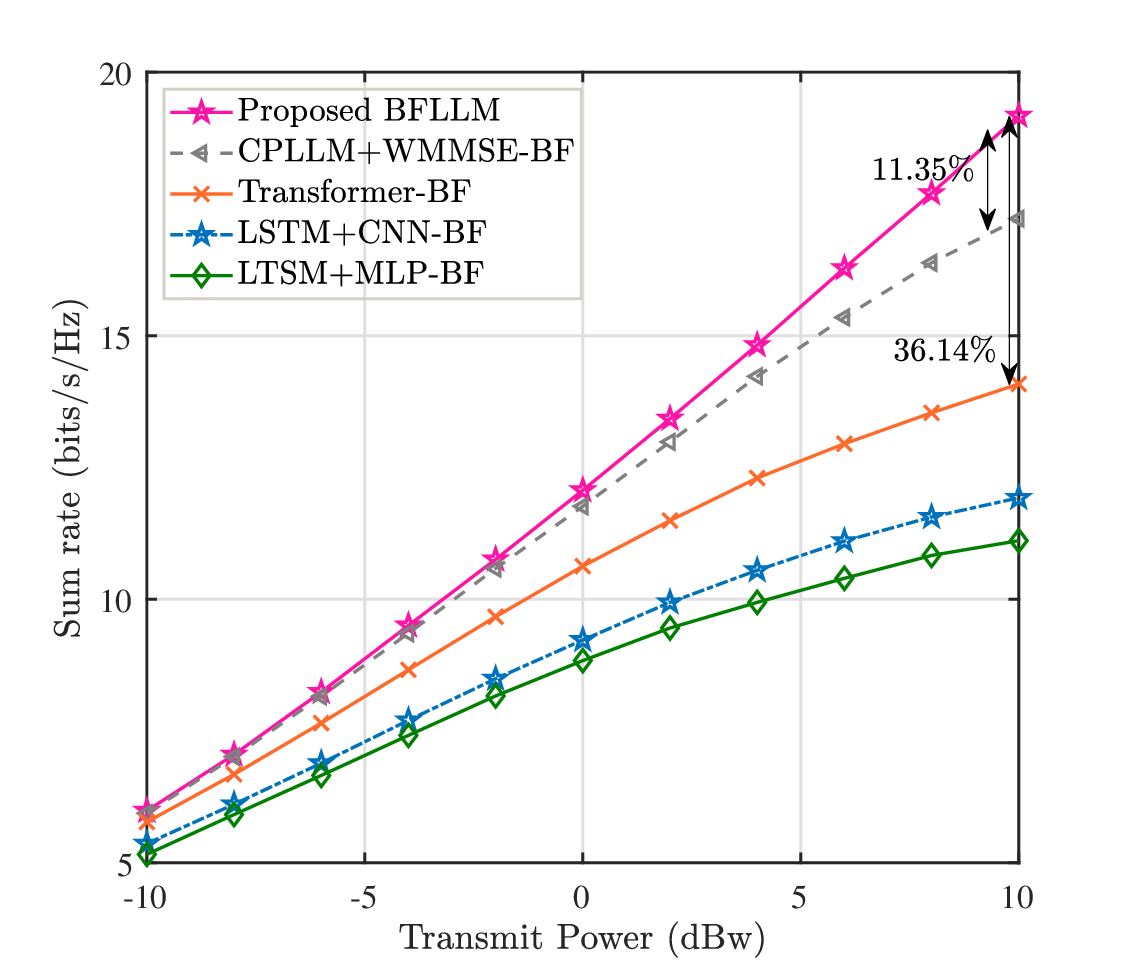}
\caption{Comparison of different predictive beamforming approaches under various transmit power $P_T$.}
\label{fig:sum_rate_power}
\end{figure}

Fig. \ref{fig:sum_rate_power} compares the proposed BFLLM approach with the benchmark methods under different transmit power levels of the LEO satellite. As the transmit power increases, the sum rate of all approaches correspondingly improves. In addition, the proposed BFLLM consistently achieves the highest sum rate across all transmit power settings. When compared to the Transformer-based predictive beamforming approach, the BFLLM method delivers a 36\% improvement in sum rate. Furthermore, compared to the CPLLM+WMMSE-BF approach, it achieves up to a 15\% increase in sum rate. While the performance of the CPLLM+WMMSE-BF approach is comparable to that of the proposed BFLLM approach at low transmit power settings, it generally incurs higher computational costs. This is because the CPLLM+WMMSE-BF approach first performs channel prediction using the proposed CPLLM, and then applies the WMMSE method to find beamforming solutions for future $T_{\rm{F}}$ slots. As a result, compared to BFLLM, the CPLLM+WMMSE-BF approach requires additional time to run $T_{\rm{F}}$ times (here $T_{\rm{F}}=4$) of the WMMSE algorithm. Moreover, this sequential process also leads to error propagation from imperfect CSI predictions and the inherent sub-optimality of WMMSE.

\begin{figure}[t]
\centering
\includegraphics[width=0.4\textwidth]{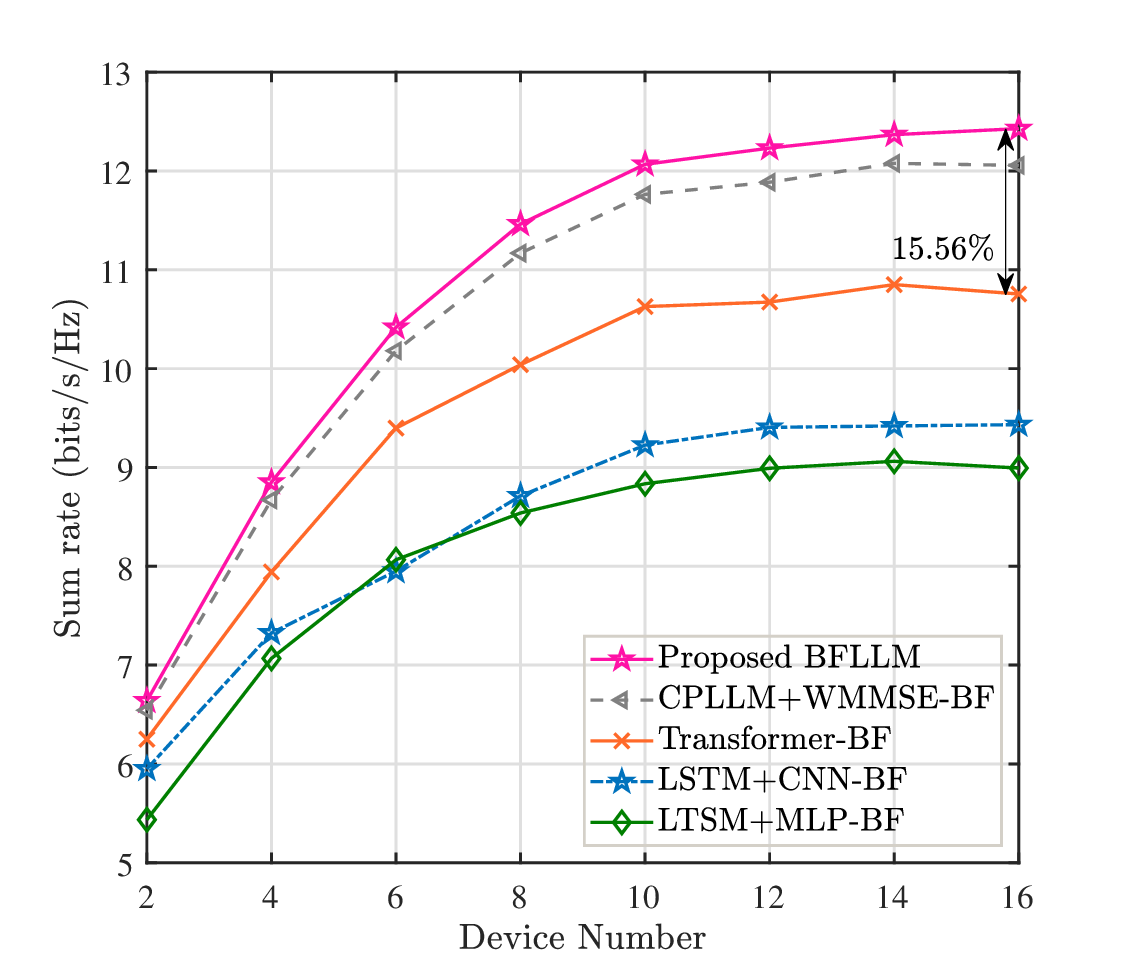}
\caption{Comparison of different predictive beamforming approaches under various device number.}
\label{fig:sum_rate_devices}
\end{figure}

Fig. \ref{fig:sum_rate_devices} compares the sum rates of all predictive beamforming approaches under different device number settings. As the number of devices increases, the sum rates of all methods initially improve but eventually saturate. This is because additional devices benefit from effectively exploiting the multiplexing gain, but the increase in device number also raises inter-device interference. Consequently, when the number of devices becomes too large, the constrained power resources are shared among more users, resulting in lower individual data rates and limiting the sum rate. In addition, the proposed approach consistently achieves the highest sum rate among all benchmarks. This performance gain is attributed to its direct prediction of beamforming strategies, which helps mitigate error propagation, as well as the impressive reasoning capabilities of the pretrained LLM.

\begin{figure}[t]
\centering
\includegraphics[width=0.4\textwidth]{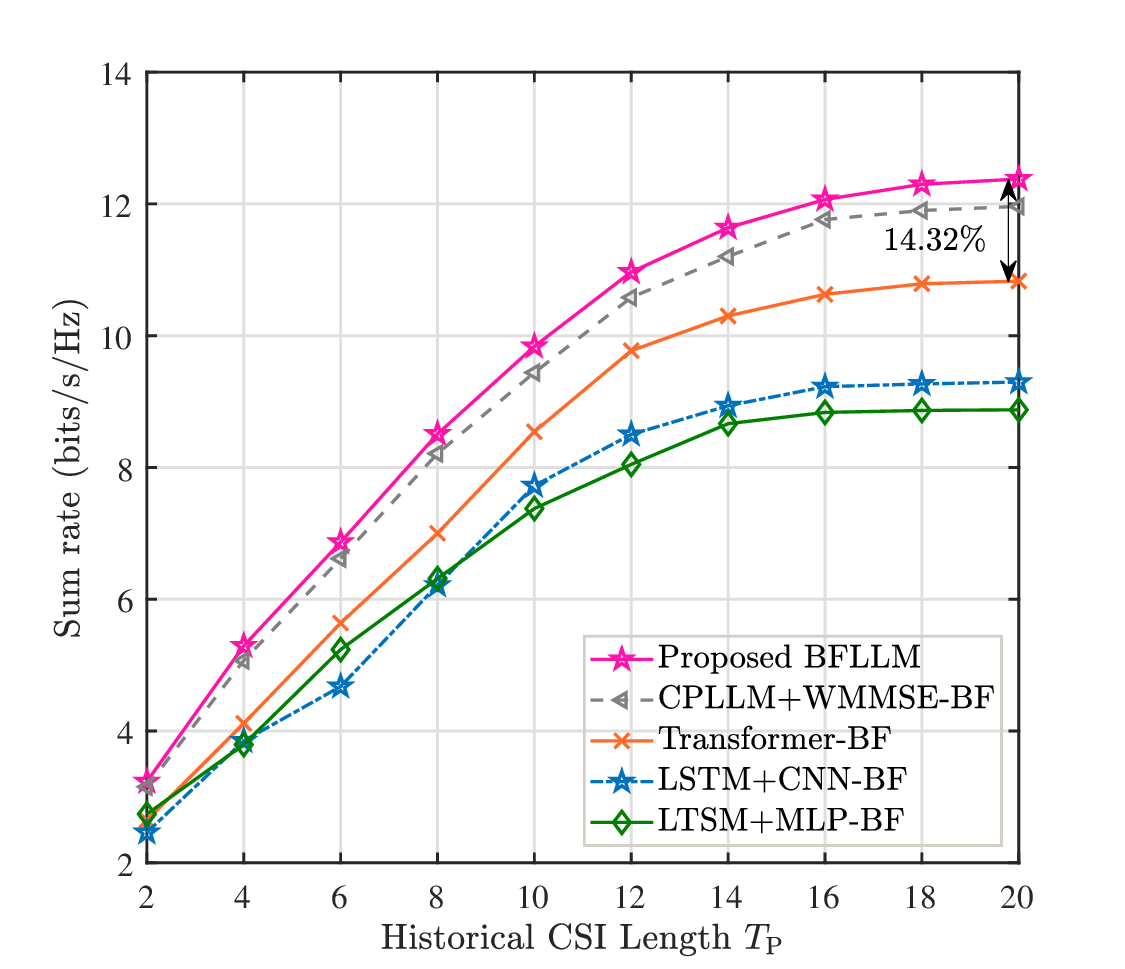}
\caption{Comparison of different predictive beamforming approaches under various historical CSI length.}
\label{fig:sum_rate_input_length}
\end{figure}

Fig. \ref{fig:sum_rate_input_length} further investigates the effect of historical CSI length on predictive beamforming performance. Consistently, our CPLLM outperforms all benchmark methods. Moreover, as the input length increases, the performance of all approaches initially improves and then gradually saturates. This is because a longer historical CSI provides richer temporal information, enabling more accurate capture of channel correlations and better prediction of beamforming strategies for future time slots. However, once the historical length reaches a certain point, the performance gain from additional data diminishes, as the models may approach their limits in leveraging temporal correlations.

\subsection{Ablation Study}
In this subsection, we present ablation studies to evaluate the role of different design components, i.e., pretrained LLM, LoRA adapter rank, and CSI decoder head, in the proposed CPLLM and BFLLM approaches.

\begin{figure}[t!]
\subfigure[]{\includegraphics[width=0.49\linewidth]{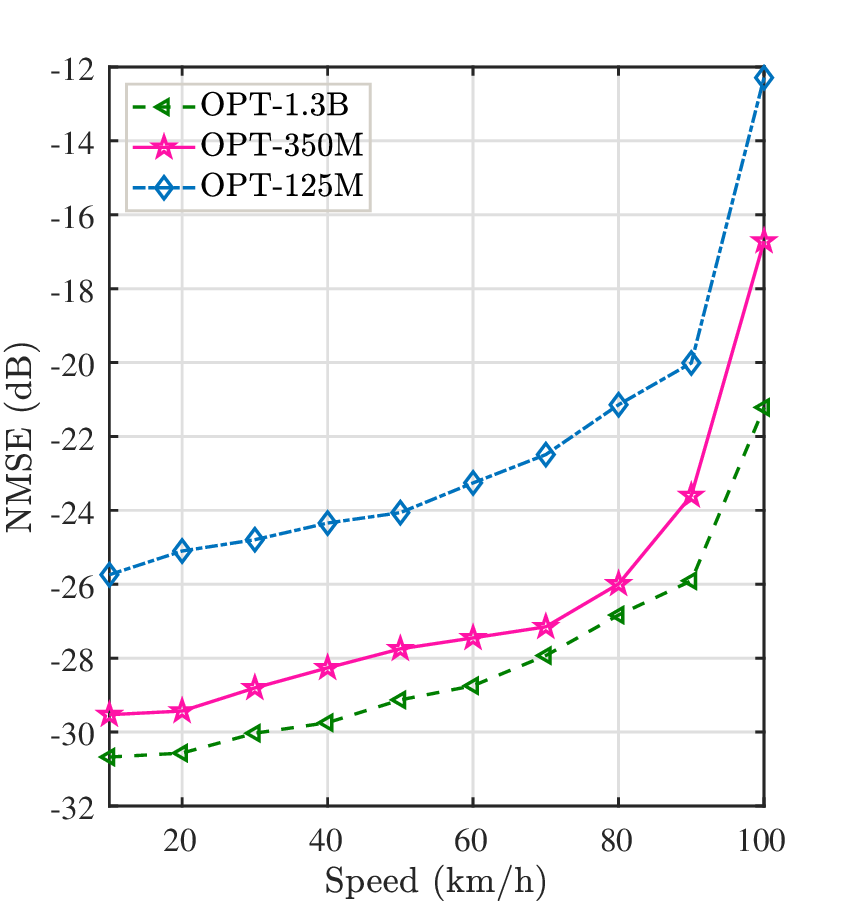}\label{fig:mnist_acc_algs_Dir0p01}}
\subfigure[]{\includegraphics[width=0.49\linewidth]{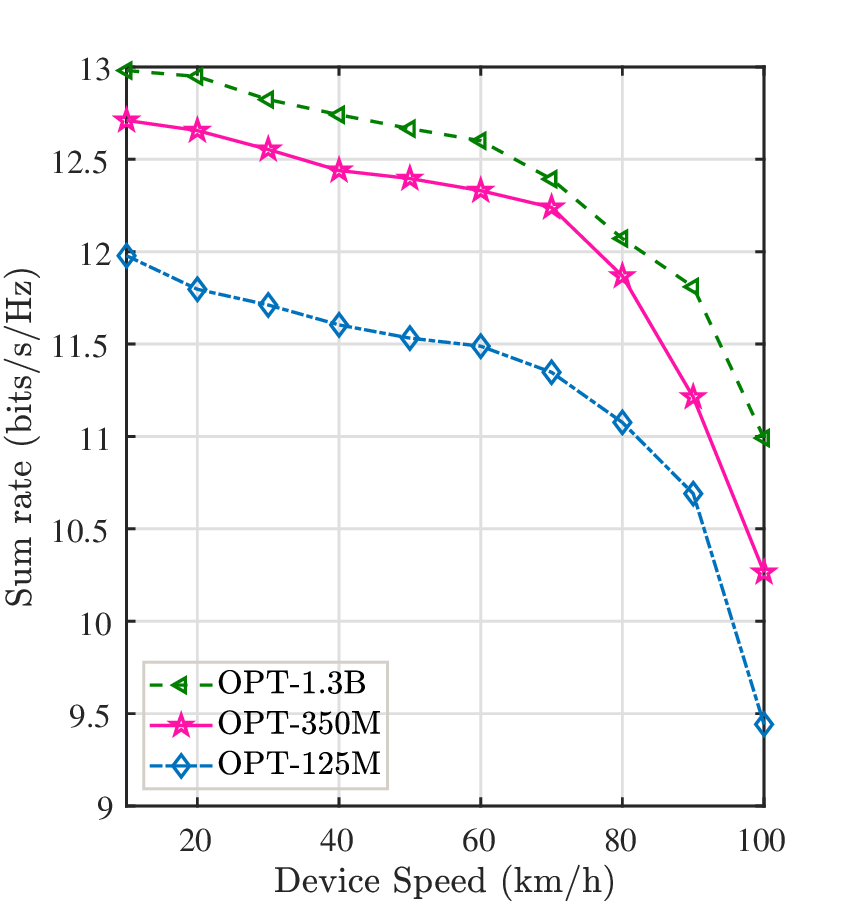}\label{fig:mnist_acc_algs_Dir0p1}}
\caption{Impact of LLM backbone on the performance of channel prediction and predictive beamforming: (a) NMSE for channel prediction (b) Sum rate for predictive beamforming.}
\label{fig:impact_LLMs}
\end{figure}

In Fig. \ref{fig:impact_LLMs}, we examine the influence of the LLM backbone on the performance of channel prediction and predictive beamforming. The LLM backbone used in this work is OPT-350M, which contains 350 million parameters. For comparison, we replace it with a smaller model, OPT-125M (125 million parameters), and a larger model, OPT-1.3B (1.3 billion parameters). For both channel prediction and predictive beamforming tasks, we can see that the performance of the LLM backbones shows: OPT-125M $<$ OPT-350M $<$ OPT-1.3B. This trend is due to the fact that as the number of parameters increases, pretrained LLMs exhibit enhanced reasoning and generalization capabilities, which in turn improves the performance of the proposed LLM-based channel prediction and predictive beamforming approaches. In addition, it is observed that the performance gap between OPT-1.3B and OPT-350M is smaller than that between OPT-350M and OPT-125M. This indicates that the performance gains from increasing the LLM backbone size may diminish as the model grows larger. Moreover, larger models also incur significantly higher computational costs. Therefore, the selection of an LLM backbone requires a careful balance between performance and computational efficiency.

\begin{table}[ht]\small
\caption{Comparison of different LoRA Adapter Rank}
\centering
\label{tab:impacts_lora_rank}
\begin{tabular}{p{0.6cm}<{\centering}|p{1.2cm}<{\centering}|p{1.4cm}<{\centering}|p{1.6cm}<{\centering}|p{1.6cm}<{\centering}}
\hline
LoRA $r$ & NMSE (dB) & Sum rate (bits/s/Hz) & Model Parameters & \% Trainable Parameters\\
\hline
0   & -17.7936 & 7.0652 &  338,215,812 &  2.07\%\\
2   & -21.3844 & 8.7965 & 338,609,028 & 2.19\% \\
4   & -23.9098 & 9.9116 & 339,002,244 & 2.30\% \\
6   & -25.6045 & 11.2128 & 339,395,460 & 2.41\% \\
8   & -26.4680 & 12.0671 & 339,788,676 & 2.52\% \\
10  & -26.7724 & 12.3962 & 340,181,892 & 2.64\% \\
12  & -26.9186& 12.9782 & 340,575,108 & 2.75\% \\
14  & -27.1463 & 13.0886 & 340,968,324 & 2.87\% \\
$d_{\rm{LLM}}$  &  -12.1635 & 6.3245 &  338,215,812 & 100\% \\
\hline
\end{tabular}
\end{table}

Table \ref{tab:impacts_lora_rank} presents the impact of the LoRA adapter rank, i.e., $r$, on the performance of both channel prediction and predictive beamforming. Note that $r=0$ indicates that all parameters in the LLM backbone are frozen with no LoRA adaptation applied, while $r=d_{\rm{LLM}}$ corresponds to full-parameter tuning, where all parameters are updated during fine-tuning. Interestingly, fine-tuning only a small subset of parameters (i.e., $0 < r < d_{\rm{LLM}}$) consistently outperforms both the fully frozen ($r=0$) and fully fine-tuned ($r=d_{\rm{LLM}}$) settings. This is because pretrained LLMs are originally optimized on textual corpora and are not inherently designed for channel prediction and predictive beamforming tasks. Moreover, full-parameter tuning may compromise the generalizable knowledge embedded within the pretrained LLM, whereas selective adaptation better preserves its transferable capabilities while improving task-specific performance. As the LoRA adapter rank $r$ increases, both the total number of model parameters and the proportion of trainable parameters grow accordingly, inevitably leading to higher computational costs. At the same time, a larger $r$ also improves the performance of channel prediction and predictive beamforming. In addition, when the LoRA adapter rank exceeds 8, the performance gains become less significant, especially when LoRA rank exceeds 12, the performance gain is merely negligible. Therefore, selecting an appropriate LoRA adapter rank requires balancing performance gains with computational efficiency.

\begin{figure}[t!]
\subfigure[]{\includegraphics[width=0.49\linewidth]{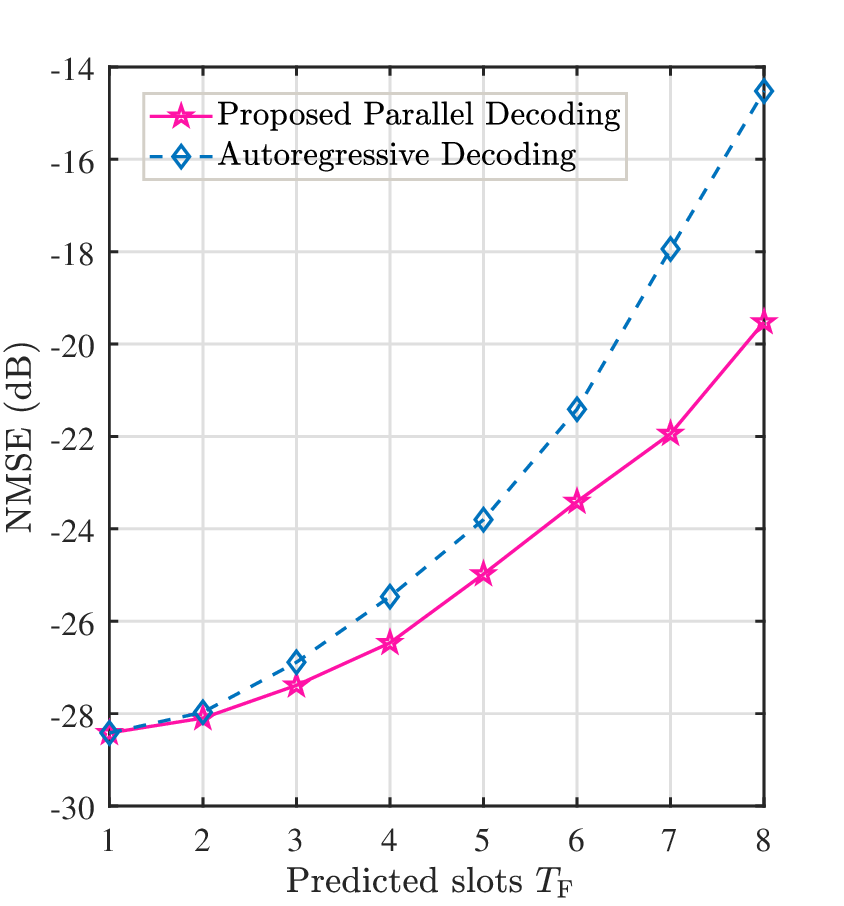}\label{fig:impact_head_CP}}
\subfigure[]{\includegraphics[width=0.49\linewidth]{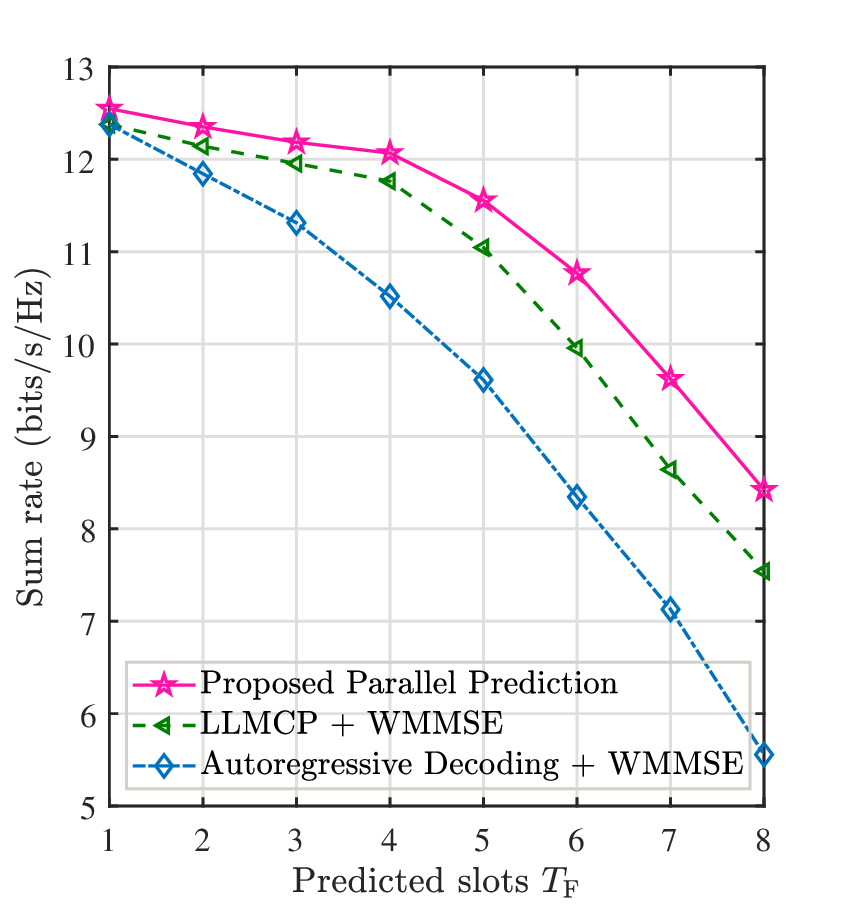}\label{fig:impact_head_BF}}
\caption{Impact of predicted slots on the performance of channel prediction and predictive beamforming: (a) NMSE for channel prediction (b) Sum rate for predictive beamforming.}
\label{fig:impact_head}
\end{figure}

In Fig. \ref{fig:impact_head},  we examine the impact of the length of predicted time slots (i.e., $T_{\rm{F}}$) on the performance of channel prediction and predictive beamforming. Fig. \ref{fig:impact_head_CP} evaluates the channel prediction performance for different values of $T_{\rm{F}}$, and further assesses the effectiveness of the proposed parallel CSI decoder by comparing it with the conventional autoregressive decoding strategy.
Specifically, for each $T_{\rm{F}}$, the proposed approach simultaneously predicts the CSI for $T_{\rm{F}}$ future slots, whereas autoregressive decoding generates the CSI for each slot sequentially, predicting one slot CSI at a time by appending the previously generated tokens to the historical CSI as the input of CPLLM. It is observed that the NMSE of both the proposed CSI decoder and the autoregressive decoder increases along with $T_{\rm{F}}$ increases, as a longer prediction length raises the complexity of the channel prediction task, thereby reducing prediction performance. However, the NMSE increases at a significantly slower rate for the proposed parallel prediction approach compared to the autoregressive decoding approach. This is because the autoregressive method generates future CSI iteratively, relying on previously predicted results, which introduces error propagation in the sequential prediction process. Additionally, autoregressive decoding requires more inference time, as it involves $T_{\rm{F}}$ forward passes through the model, whereas the proposed approach only requires a single forward pass.
Fig. \ref{fig:impact_head_BF} evaluates the effect of $T_{\rm{F}}$ on the predictive beamforming performance of the proposed BFLLM. Since BFLLM directly outputs beamforming strategies for future slots, it is incompatible with the autoregressive decoding approach. For comparison, we compare the proposed parallel predictive beamforming decoder with two alternatives: 1) the CPLLM+WMMSE approach described in Section \ref{subsec:simu_BF}, and 2) the autoregressive channel prediction method followed by the WMMSE method, which sequentially predicts the CSI for one future slot then finding beamforming solution by WMMSE, and repeats this process over $T_{\rm{F}}$ slots. As $T_{\rm{F}}$ increases, the predictive beamforming performance deteriorates due to the growing complexity of the task, with the performance drop becoming more noticeable when $T_{\rm{F}}$ exceeds 5. Therefore, selecting an appropriate $T_{\rm{F}}$ is essential for balancing performance while mitigating the channel ageing problem.

\section{Conclusion}\label{sec:conclusion}
In this work, we propose an LLM-based channel prediction framework, i.e., CPLLM, to tackle the channel aging problem in LEO satellite communications by predicting future CSI from historical CSI data. To leverage the powerful reasoning and generalization capabilities of pretrained LLMs, we first design a CSI encoder that aligns CSI data with the textual embeddings used in the LLM. A CSI decoder module is then introduced to simultaneously generate CSI for multiple future time slots. To enhance prediction performance, we adopt LoRA for parameter-efficient fine-tuning of the proposed framework. Furthermore, we extend the proposed CPLLM to directly generate beamforming solutions for future slots based on historical CSI, i.e., BFLLM, thereby reducing signal processing overhead in LEO satellite communications. Simulation results demonstrate the superiority of the proposed methods over existing LLM-based channel predictors and deep learning-based predictive beamforming techniques.

\bibliographystyle{IEEEtran}
\bibliography{IEEEabrv,cited}

\begin{thebibliography}{10}
\providecommand{\url}[1]{#1}
\csname url@rmstyle\endcsname
\providecommand{\newblock}{\relax}
\providecommand{\bibinfo}[2]{#2}
\providecommand\BIBentrySTDinterwordspacing{\spaceskip=0pt\relax}
\providecommand\BIBentryALTinterwordstretchfactor{4}
\providecommand\BIBentryALTinterwordspacing{\spaceskip=\fontdimen2\font plus
\BIBentryALTinterwordstretchfactor\fontdimen3\font minus
  \fontdimen4\font\relax}
\providecommand\BIBforeignlanguage[2]{{%
\expandafter\ifx\csname l@#1\endcsname\relax
\typeout{** WARNING: IEEEtran.bst: No hyphenation pattern has been}%
\typeout{** loaded for the language `#1'. Using the pattern for}%
\typeout{** the default language instead.}%
\else
\language=\csname l@#1\endcsname
\fi
#2}}

\bibitem{10399870}
X.~Luo, H.-H. Chen, and Q.~Guo, ``{LEO}/{VLEO} satellite communications in 6{G}
  and beyond networks-technologies, applications, and challenges,'' \emph{IEEE
  Network}, vol.~38, no.~5, pp. 273--285, 2024.

\bibitem{9982369}
T.~Ma, B.~Qian, X.~Qin, X.~Liu, H.~Zhou, and L.~Zhao, ``Satellite-terrestrial
  integrated 6{G}: An ultra-dense {LEO} networking management architecture,''
  \emph{IEEE Wireless Commun.}, vol.~31, no.~1, pp. 62--69, 2024.

\bibitem{8654189}
R.~T. Schwarz, T.~Delamotte, K.-U. Storek, and A.~Knopp, ``{MIMO} applications
  for multibeam satellites,'' \emph{IEEE Trans. Broadcasting}, vol.~65, no.~4,
  pp. 664--681, 2019.

\bibitem{9110855}
L.~You, K.-X. Li, J.~Wang, X.~Gao, X.-G. Xia, and B.~Ottersten, ``Massive
  {MIMO} transmission for {LEO} satellite communications,'' \emph{IEEE J. Sel.
  Areas Commun.}, vol.~38, no.~8, pp. 1851--1865, 2020.

\bibitem{10946972}
Z.~Chen, H.~Shin, and A.~Nallanathan, ``Generative diffusion model-based
  variational inference for mimo channel estimation,'' \emph{IEEE Trans.
  Commun.}, pp. 1--1, 2025.

\bibitem{1512123}
K.~Baddour and N.~Beaulieu, ``Autoregressive modeling for fading channel
  simulation,'' \emph{IEEE Trans. Wireless Commun.}, vol.~4, no.~4, pp.
  1650--1662, 2005.

\bibitem{9127447}
H.~Yin, H.~Wang, Y.~Liu, and D.~Gesbert, ``Addressing the curse of mobility in
  massive {MIMO} with prony-based angular-delay domain channel predictions,''
  \emph{IEEE J. Sel. Areas Commun.}, vol.~38, no.~12, pp. 2903--2917, 2020.

\bibitem{9210016}
H.~Kim, S.~Kim, H.~Lee, C.~Jang, Y.~Choi, and J.~Choi, ``Massive {MIMO} channel
  prediction: Kalman filtering vs. machine learning,'' \emph{IEEE Trans.
  Commun.}, vol.~69, no.~1, pp. 518--528, 2021.

\bibitem{8813020}
W.~Jiang and H.~D. Schotten, ``Neural network-based fading channel prediction:
  A comprehensive overview,'' \emph{IEEE Access}, vol.~7, pp.
  118\,112--118\,124, 2019.

\bibitem{8801923}
Y.~Zhu, X.~Dong, and T.~Lu, ``An adaptive and parameter-free recurrent neural
  structure for wireless channel prediction,'' \emph{IEEE Trans. Commun.},
  vol.~67, no.~11, pp. 8086--8096, 2019.

\bibitem{9439942}
Y.~Zhang, Y.~Wu, A.~Liu, X.~Xia, T.~Pan, and X.~Liu, ``Deep learning-based
  channel prediction for {LEO} satellite massive {MIMO} communication system,''
  \emph{IEEE Wireless Commun. Letters}, vol.~10, no.~8, pp. 1835--1839, 2021.

\bibitem{10089512}
I.~Helmy, P.~Tarafder, and W.~Choi, ``Lstm-gru model-based channel prediction
  for one-bit massive {MIMO} system,'' \emph{IEEE Trans. Veh. Technol.},
  vol.~72, no.~8, pp. 11\,053--11\,057, 2023.

\bibitem{9921297}
G.~Liu, Z.~Hu, L.~Wang, J.~Xue, H.~Yin, and D.~Gesbert, ``Spatio-temporal
  neural network for channel prediction in massive {MIMO}-{OFDM} systems,''
  \emph{IEEE Trans. Commun.}, vol.~70, no.~12, pp. 8003--8016, 2022.

\bibitem{10582829}
B.~Liu, X.~Liu, S.~Gao, X.~Cheng, and L.~Yang, ``{LLM4CP}: Adapting large
  language models for channel prediction,'' \emph{J. Commun. and Infor. Netw.},
  vol.~9, no.~2, pp. 113--125, 2024.

\bibitem{4599181}
A.~Wiesel, Y.~C. Eldar, and S.~Shamai, ``Zero-forcing precoding and generalized
  inverses,'' \emph{IEEE Trans. Signal Processing}, vol.~56, no.~9, pp.
  4409--4418, 2008.

\bibitem{4712693}
S.~S. Christensen, R.~Agarwal, E.~De~Carvalho, and J.~M. Cioffi, ``Weighted
  sum-rate maximization using weighted {MMSE} for {MIMO}-{BC} beamforming
  design,'' \emph{IEEE Trans. Wireless Commun.}, vol.~7, no.~12, pp.
  4792--4799, 2008.

\bibitem{10440321}
Z.~Xiang, X.~Gao, K.-X. Li, and X.-G. Xia, ``Massive {MIMO} downlink
  transmission for multiple {LEO} satellite communication,'' \emph{IEEE Trans.
  Commun.}, vol.~72, no.~6, pp. 3352--3364, 2024.

\bibitem{9328170}
I.~Ahmad, K.~D. Nguyen, N.~Letzepis, G.~Lechner, and V.~Joroughi,
  ``Zero-forcing precoding with partial csi in multibeam high throughput
  satellite systems,'' \emph{IEEE Trans. Veh. Technol.}, vol.~70, no.~2, pp.
  1410--1420, 2021.

\bibitem{9165811}
J.~Chu, X.~Chen, C.~Zhong, and Z.~Zhang, ``Robust design for {NOMA}-based
  multibeam {LEO} satellite internet of things,'' \emph{IEEE Int. Things J.},
  vol.~8, no.~3, pp. 1959--1970, 2021.

\bibitem{9694506}
L.~You, X.~Qiang, K.-X. Li, C.~G. Tsinos, W.~Wang, X.~Gao, and B.~Ottersten,
  ``Hybrid analog/digital precoding for downlink massive {MIMO} {LEO} satellite
  communications,'' \emph{IEEE Trans. Wireless Commun.}, vol.~21, no.~8, pp.
  5962--5976, 2022.

\bibitem{9000850}
W.~Ma, C.~Qi, Z.~Zhang, and J.~Cheng, ``Sparse channel estimation and hybrid
  precoding using deep learning for millimeter wave massive {MIMO},''
  \emph{IEEE Trans. Commun.}, vol.~68, no.~5, pp. 2838--2849, 2020.

\bibitem{10550141}
M.~Ying, X.~Chen, Q.~Qi, and W.~Gerstacker, ``Deep learning-based joint channel
  prediction and multibeam precoding for {LEO} satellite internet of things,''
  \emph{IEEE Trans. Wireless Commun.}, vol.~23, no.~10, pp. 13\,946--13\,960,
  2024.

\bibitem{9143143}
W.~Yuan, C.~Liu, F.~Liu, S.~Li, and D.~W.~K. Ng, ``Learning-based predictive
  beamforming for {UAV} communications with jittering,'' \emph{IEEE Wireless
  Commun. Letters}, vol.~9, no.~11, pp. 1970--1974, 2020.

\bibitem{10138552}
C.~Liu, S.~Li, W.~Yuan, X.~Liu, and D.~W.~K. Ng, ``Predictive precoder design
  for {OTFS}-enabled {URLLC}: A deep learning approach,'' \emph{IEEE J. Sel.
  Areas Commun.}, vol.~41, no.~7, pp. 2245--2260, 2023.

\bibitem{10741218}
S.~Zhang, S.~Zhang, W.~Yuan, and T.~Q.~S. Quek, ``Transformer-empowered
  predictive beamforming for rate-splitting multiple access in non-terrestrial
  networks,'' \emph{IEEE Trans. Wireless Commun.}, vol.~23, no.~12, pp.
  19\,776--19\,788, 2024.

\bibitem{6395846}
E.~T. Michailidis, P.~Theofilakos, and A.~G. Kanatas, ``Three-dimensional
  modeling and simulation of {MIMO} mobile-to-mobile via stratospheric relay
  fading channels,'' \emph{IEEE Trans. Veh. Technol.}, vol.~62, no.~5, pp.
  2014--2030, 2013.

\bibitem{10742081}
J.~Wang, S.~Gong, J.~Xiao, P.~Guo, J.~Wang, W.~Xie, and X.~Li, ``A lightweight
  channel prediction network for {UAV}-{LEO} satellite communications,''
  \emph{IEEE Wireless Commun. Letters}, vol.~14, no.~1, pp. 113--117, 2025.

\bibitem{662636}
I.~Ali, N.~Al-Dhahir, and J.~Hershey, ``Doppler characterization for {LEO}
  satellites,'' \emph{IEEE Trans. Commun.}, vol.~46, no.~3, pp. 309--313, 1998.

\bibitem{dosovitskiy2020image}
A.~Dosovitskiy, L.~Beyer, A.~Kolesnikov, D.~Weissenborn, X.~Zhai,
  T.~Unterthiner, M.~Dehghani, M.~Minderer, G.~Heigold, S.~Gelly,
  \emph{et~al.}, ``An image is worth 16x16 words: Transformers for image
  recognition at scale,'' \emph{arXiv preprint arXiv:2010.11929}, 2020.

\bibitem{takase2019positional}
S.~Takase and N.~Okazaki, ``Positional encoding to control output sequence
  length,'' \emph{arXiv preprint arXiv:1904.07418}, 2019.

\bibitem{zhang2022opt}
S.~Zhang, S.~Roller, N.~Goyal, M.~Artetxe, M.~Chen, S.~Chen, C.~Dewan, M.~Diab,
  X.~Li, X.~V. Lin, \emph{et~al.}, ``Opt: Open pre-trained transformer language
  models,'' \emph{arXiv preprint arXiv:2205.01068}, 2022.

\bibitem{touvron2023llama}
H.~Touvron, T.~Lavril, G.~Izacard, X.~Martinet, M.-A. Lachaux, T.~Lacroix,
  B.~Rozi{\`e}re, N.~Goyal, E.~Hambro, F.~Azhar, \emph{et~al.}, ``Llama: Open
  and efficient foundation language models,'' \emph{arXiv preprint
  arXiv:2302.13971}, 2023.

\bibitem{NEURIPS2023_6dcf277e}
H.~Liu, C.~Li, Q.~Wu, and Y.~J. Lee, ``Visual instruction tuning,'' in
  \emph{Advances in Neural Information Processing Systems}.\hskip 1em plus
  0.5em minus 0.4em\relax Curran Associates, Inc., pp. 34\,892--34\,916.

\bibitem{hu2022lora}
E.~J. Hu, Y.~Shen, P.~Wallis, Z.~Allen-Zhu, Y.~Li, S.~Wang, L.~Wang, and
  W.~Chen, ``Lo{RA}: Low-rank adaptation of large language models,'' in
  \emph{International Conference on Learning Representations}, 2022.

\bibitem{10021887}
J.~Zhang, G.~Zheng, Y.~Zhang, I.~Krikidis, and K.-K. Wong, ``Deep learning
  based predictive beamforming design,'' \emph{IEEE Trans. Veh. Technol.},
  vol.~72, no.~6, pp. 8122--8127, 2023.

\bibitem{9832933}
H.~Jiang, M.~Cui, D.~W.~K. Ng, and L.~Dai, ``Accurate channel prediction based
  on transformer: Making mobility negligible,'' \emph{IEEE J. Sel. Areas
  Commun.}, vol.~40, no.~9, pp. 2717--2732, 2022.

\end{thebibliography}
\end{document}